\documentclass[12pt]{article}

\usepackage{amsmath,amssymb,amsfonts,graphicx,cite,color,feynarts,soul,axodraw}
\usepackage{epsfig}
\usepackage{mciteplus}
\usepackage{wasysym}
\usepackage{multirow}
\usepackage{upgreek}
\usepackage{psfrag}
\usepackage{url}
\usepackage{rotate}
\usepackage[normalem]{ulem}
\usepackage{tikz}

\input{paperdef}

\graphicspath{{figs/}}

\evensidemargin \oddsidemargin
\allowdisplaybreaks

\newcommand{\HiggsBounds}{{\tt HiggsBounds}}

\newcommand{\GeV}{{\;\mathrm{GeV}}}

\renewenvironment{thebibliography}[1]{%
\begin{oldthebibliography}{#1}%
\small%
\raggedright%
\setlength{\itemsep}{5pt}%
}%
{%
\end{oldthebibliography}%
}

\newcommand{\MHexp}{125.6}

\begin{document}
\thispagestyle{empty}

\def\thefootnote{\fnsymbol{footnote}}

\hfill {\tt DESY 13-015}

\hfill {\tt MPP-2013-291}

\vspace{0.5cm}

\begin{center}
{\large\sc {\bf Implications of LHC search results \\[.5em]
on the \boldmath{$W$} boson mass prediction in the MSSM}}

\vspace{1cm}

{\sc
 S.~Heinemeyer$^{1}$%
\footnote{email: Sven.Heinemeyer@cern.ch}%
, W.~Hollik$^{2}$%
\footnote{email: hollik@mpp.mpg.de}%
, G.~Weiglein$^{3}$%
\footnote{email: Georg.Weiglein@desy.de}%
,~and L.~Zeune$^{3}$%
\footnote{email: Lisa.Zeune@desy.de}
}

\vspace*{.7cm}

{\sl
$^1$Instituto de F\'isica de Cantabria (CSIC-UC), Santander,  Spain

\vspace*{0.1cm}
$^2$Max-Planck-Institut f\"ur Physik (Werner-Heisenberg-Institut), 
F\"ohringer Ring 6, 
D--80805 M\"unchen, Germany

\vspace*{0.1cm}
$^3$DESY, Notkestra\ss e 85, D--22607 Hamburg, Germany
}

\end{center}

\vspace*{0.1cm}

\begin{abstract}
\noindent
We present the currently most precise $W$~boson mass ($\MW$) prediction in the 
Minimal Supersymmetric Standard Model (MSSM) and discuss how it is affected by 
recent results from the LHC. The evaluation includes the full one-loop result 
and all known higher order corrections of SM and SUSY type.
We show the MSSM prediction in the $\MW$--$\mt$ plane, 
taking into account constraints from Higgs and SUSY searches. 
We point out
that even if stops and sbottoms are heavy, relatively large SUSY
contributions  
to $\MW$ are possible if either charginos, neutralinos or sleptons are light.
In particular we analyze the effect on the $\MW$ prediction of the Higgs
signal at about $\MHexp \gev$, which within the MSSM can in
principle be interpreted
as the light or the heavy $\cp$-even Higgs boson.
For both interpretations the predicted MSSM region for $\MW$ is in good
agreement with the experimental measurement.
We furthermore discuss the impact of possible future LHC results 
in the stop sector on the $\MW$ prediction, considering both the cases
of improved limits and of the detection of a scalar top quark.
\\ 
\end{abstract}

\def\thefootnote{\arabic{footnote}}
\setcounter{page}{0}
\setcounter{footnote}{0}

\newpage


\section{Introduction}
\label{sect:Introduction}

The recent discovery of a signal with a mass of around
$\MHexp \gev$ in the Higgs searches at
ATLAS~\cite{atlas:2012gk} and CMS~\cite{cms:2012gu} 
is compatible with the Higgs boson postulated by the Standard Model (SM), 
but it can also be interpreted in a variety of models of physics beyond
the SM. On the other hand, the direct searches for physics beyond the SM 
have not resulted in a signal so far. In order to enhance the
sensitivity for discriminating between different models of the underlying
physics, it is useful to complement the measurements of the properties of
the new state with other high-precision observables that have
sensitivity to the quantum level, i.e.\ to loop contributions involving
in principle all the particles of the considered model. 

In this context, the relation between the $W$ boson mass, $\MW$,
and the 
$Z$ boson mass, $\MZ$, in terms of the fine-structure constant, $\al$, the
Fermi constant, $G_{\mu}$, and the parameters entering via loop
contributions plays a crucial role. The accuracy of the measurement of
the $W$ boson mass has significantly been improved with the latest results
presented by CDF~\cite{Aaltonen:2012bp} and D\O~\cite{Abazov:2012bv}.
Together with the results obtained at LEP~\cite{Alcaraz:2006mx} this gives rise
to the latest world average of~\cite{Group:2012gb,ALEPH:2005ab}
\begin{align}
\MW^{\rm exp} = 80.385 \pm 0.015 \gev,
\end{align}
i.e.\ to a relative experimental accuracy of better than $2 \times 10^{-4}$. 
Furthermore, the improved measurement of the top-quark mass, $\mt$, at
the Tevatron and the LHC (see below for a discussion of the physical
interpretation of those measurements) has improved the accuracy of the
theoretical prediction for $\MW$, since the experimental error of the
input parameter $\mt$ constitutes a dominant source of (parametric)
uncertainty in the theoretical prediction, see e.g.\
\citere{Heinemeyer:2003ud}. Further observables that have a high
sensitivity for testing electroweak physics at the quantum level are in 
particular the effective leptonic weak mixing angle at the $Z$-resonance,
$\sweff$, the anomalous magnetic moment of the muon, $(g-2)_\mu$, and rare
$b$ decays such as $b \to s \ga$. The interpretation of the constraints
from $\sweff$ are complicated by the fact that the two single most
precise measurements, $A_{\rm LR}$ by SLD~\cite{ALEPH:2005ab} and 
$A_b^{\rm FB}$ at LEP~\cite{ALEPH:2005ab}, differ from each other by
more than $3\,\si$, see e.g.\ \citere{Heinemeyer:2010yp} for a recent
discussion. While the experimental value of $(g-2)_\mu$ shows a
significant deviation from the SM prediction at the level of 3--4$\,\si$,
which led to many interpretations in terms of new physics models
(see e.g.\ \citeres{Stockinger:2006zn,Miller:2012opa,Jegerlehner:2009ry} 
for reviews), the analysis of rare $b$ decays
so far has 
been inconclusive~\cite{Amhis:2012bh}. 

We will concentrate in the following on the prediction for the $W$ boson
mass and, taking into account the latest experimental results,
compare the prediction of the SM with that of its most popular
extension, the Minimal Supersymmetric Standard Model 
(MSSM)~\cite{Nilles:1983ge,Haber:1984rc,Barbieri:1987xf}. 
Within the SM, the interpretation of the discovered new state as the 
SM Higgs boson implies that there is no unknown parameter anymore in the
prediction for $\MW$. This fact considerably sharpens both the comparison
with the experimental result for $\MW$ and with predictions in
extensions of the SM such as the MSSM. 
Our analysis within the MSSM updates previous studies, see in particular 
\citeres{Heinemeyer:2006px,Heinemeyer:2004gx} and references therein. 
Our results are based on the currently most precise prediction for $\MW$ 
in the MSSM, which we compare with the result in the SM. The MSSM prediction
consists of a complete
one-loop calculation for the general case of complex
parameters (without flavor violation in the sfermion
sector~\cite{Heinemeyer:2004by}), combined with all known higher-order
corrections of SM and supersymmetric (SUSY)
type. Compared to the result employed in
\citere{Heinemeyer:2006px}, the MSSM prediction used in the present
analysis has been improved in several respects: the one-loop result in
the MSSM has been reevaluated and coded in a more flexible way, which
permits an improved treatment of regions of parameter space that can
lead to numerical instabilities and furthermore provides the
functionality to easily implement results for non-minimal SUSY models
(see \citere{MWnmssm} and also \citere{Domingo:2011uf} for the case of the NMSSM); the incorporation of
the state-of-the-art SM result has been improved using the expressions
given in \citere{Awramik:2006uz}.

The top quark mass used in our evaluation corresponds to the pole mass.
In our results it could easily be re-expressed in terms of a properly
defined short distance mass such as the \msbar\ or \drbar\ mass.
The parameter measured with high precision via direct reconstruction
at the Tevatron and the LHC is expected to be close to the top pole
mass, and we adopt this interpretation in the following. For a
discussion of the systematic uncertainties arising from the difficulties
how to relate the measured mass parameter to the pole mass see
\citeres{Skands:2007zg,Hoang:2008xm}.  

Extensive searches for SUSY particles have been performed by ATLAS and CMS.
No supersymmetric particles have been detected so far
in direct searches, and stringent 
limits were set in particular 
on the gluino mass and the mass of the squarks of the first 
two generations~\cite{SUSYsusy1,SUSYsusy2,AtlasSusy,CMSSusy},
see however
\citeres{LeCompte:2011cn,Mahbubani:2012qq}.
Substantially weaker limits have been reported for the particles of the other 
MSSM sectors, so that third-generation squarks, stops and sbottoms, 
as well as the uncolored SUSY particles are significantly 
less constrained by LHC searches, and LEP limits still give relevant
constraints~\cite{Beringer:1900zz}.

In this paper we analyze the prediction for $\MW$ in view of the
discovery of a signal in the Higgs searches at ATLAS and CMS.
Within the framework of the MSSM the lighter $\cp$-even Higgs boson 
can have
a mass of about $\MHexp \gev$ for sufficiently large $\MA$ and
sufficiently large higher-order corrections from the scalar top sector.
It is interesting to note that a mass value as high
as about $\MHexp \gev$ for the lighter $\cp$-even Higgs boson of the
MSSM implies that $\MA$ has to be in the decoupling region, $\MA \gg
\MZ$, which in turn has the consequence that the state at about $\MHexp
\gev$ has a SM-like behavior, see e.g.\ the discussion in
\citeres{Heinemeyer:2011aa,Carena:2013qia}.
However, also the interpretation of the
discovered particle as the heavy $\cp$-even Higgs state of the MSSM
is, at least in principle, a viable possibility,
see~\citeres{Heinemeyer:2011aa,Bechtle:2012jw,Bottino:2011xv,Drees:2012fb,Hagiwara:2012mga,Arbey:2012dq,Carena:2013qia}%
\footnote{This scenario is challenged by the recent ATLAS bound on
light charged Higgs bosons~\cite{AtlaschargedhiggsSUSY13}.}.
We take into account the information from the mass measurement of the
observed Higgs boson for these two cases, and for the light Higgs
interpretation we investigate the correlation between $\MW$ and 
$\Ga(h \to \ga\ga)$. The limits from Higgs searches at LEP, the Tevatron and
the LHC are incorporated with the help of the code 
{\tt HiggsBounds} (version 4.0.0)~\cite{Bechtle:2013gu,Bechtle:2011sb,Bechtle:2008jh}%
\footnote{The latest ATLAS results on light charged Higgs boson
searches~\cite{AtlaschargedhiggsSUSY13} are not included in this {\tt HiggsBounds} version (while finalizing this paper a new {\tt HiggsBounds} version including this result became available~\cite{Bechtle:2013wla}).}.
We perform scans over the relevant SUSY parameters and we 
analyze in detail the impact of different SUSY sectors on the prediction of
$\MW$. We also investigate possible effects of either future limits from 
SUSY searches at the LHC or of the detection of a scalar top quark.

This paper is organized as follows: 
In the next section we give a short summary of the relevant MSSM
sectors and specify our notation. In~\refse{sect:mwdet} and
\ref{sect:deltar} we describe the evaluation of $\MW$ in the MSSM. 
In~\refse{sect:numerics} we
present the result for $\MW$ from a global scan over the MSSM parameter
space. We investigate the contributions from all relevant MSSM 
particle sectors and analyze the impact of the observed Higgs signal
as well as from
limits arising from searches for Higgs bosons and SUSY particles. 
Effects of possible future results from SUSY searches at the LHC
are also discussed in this context.
The conclusions can be found
in~\refse{sect:conclusions}. 


\section{Particle sectors of the MSSM}
\label{sec:mssm}

The prediction for $\MW$ in the MSSM depends on the masses, 
mixing angles and
couplings of all MSSM particles.
Sfermions, charginos, neutralinos and the MSSM Higgs bosons enter 
already at the one-loop level and can give substantial contributions to $\MW$. 
In this section we briefly describe the relevant MSSM sectors and 
fix our notation for the MSSM parameters.
In our numerical analysis below we will focus on the case of real
MSSM parameters. For a discussion of the possible impact of non-zero
phases of the MSSM parameters see 
\citere{Heinemeyer:2006px}.

Contrary to the SM, two Higgs doublets are required in the MSSM, resulting in five
physical Higgs boson degrees of freedom. 
At the tree level, where possible $\cp$-violating contributions of the 
soft supersymmetry-breaking terms do not enter,
these are the light and heavy $\cp$-even Higgs bosons, $h$ and $H$, the
$\cp$-odd Higgs boson, $A$, and the charged Higgs bosons, $H^\pm$.
At lowest order the MSSM Higgs sector is fully described by 
$\MZ$ and two MSSM parameters, 
often chosen as the $\cp$-odd Higgs boson mass, $\MA$, and 
$\tb \equiv v_2/v_1$, the ratio of the two vacuum expectation values.
Higher-order corrections to the Higgs boson masses can be sizeable and must be
included. Particularly important are the one- and two-loop contributions
from top quarks and squarks. 
Accordingly, the masses of the $\cp$-even neutral Higgs bosons and the
charged Higgs boson are not free parameters (as the Higgs mass in the SM),
but can be predicted in terms of the other MSSM parameters (introduced
below).

The sfermion mass matrix in the gauge-eigenstate basis ($\sfl$, $\sfr$) 
for one generation and flavor $f$ is given by
\begin{align}
\matr{M}_{\Sf} =
\begin{pmatrix}
        M_{\sfl}^2 + \mf^2 + \MZ^2 \CZb (I_3^f - Q_f \sw^2) & \mf \; X_f \\
        \mf \; X_f    & M_{\sfr}^2 + \mf^2 + \MZ^2 \CZb Q_f \sw^2
\end{pmatrix}~.
\label{squarkmassmatrix}
\end{align}
Here $m_f$ denotes the corresponding fermion mass, 
$I_3$ is the third component of the weak isospin, 
$Q_f$ the electric charge and $\sw$ is the sine of the weak mixing angle.
The $L$--$R$ mixing of the sfermions is determined by the off-diagonal entries
\begin{align}
\mf X_f &= \mf (A_f - \mu\, \{ \CTb, \tb \}) ,
\label{squarksoftSUSYbreaking}
\end{align}
where $\CTb$ refers to up-type sfermions and $\tb$ to down-type sfermions. 
$A_f$ denotes the trilinear Higgs-sfermion coupling and $\mu$ the
Higgsino mass parameter.
The SUSY-breaking parameters are:
\begin{align}
M_{\sfl}&=
\begin{cases}
       M_{\tilde{Q}_i} & \text{for left-handed squarks} \\
       M_{\tilde{L}_i} & \text{for left-handed sleptons}
 \end{cases}\\
M_{\sfr}&=
\begin{cases}
       M_{\tilde{U}_i} & \text{for right-handed u-type squarks} \\
       M_{\tilde{D}_i} & \text{for right-handed d-type squarks}\\
       M_{\tilde{E}_i} & \text{for right-handed charged sleptons}\;,
 \end{cases}
\label{eq:ssbp}
\end{align}
where $i=1,2,3$ is the family index.
Flavor violation in the sfermion sector is neglected here (see
\citeres{Heinemeyer:2004by,AranaCatania:2011ak} for a discussion of this
kind of effects in the one-loop contributions to $\MW$).
The charged gauginos and Higgsinos mix with each other,
yielding charginos $\tilde{\chi}^{\pm}_{1,2}$. The corresponding
mass matrix is given by
\begin{equation}
\matr{M}_{\cha{}}=\left(\begin{array}{cc} M_2 & \sqrt{2} \MW \sin \beta
  \\ \sqrt{2} \MW \cos \beta & \mu \end{array}\right)\;,
\label{charginomass1}
\end{equation}
with the soft breaking parameter $M_2$.
The neutralinos are mixtures of the neutral gauginos and Higgsinos.
The neutralino mass matrix in the basis $(\tilde{B}, \tilde{W}^0, \tilde{H}_1^0, \tilde{H}_2^0)$
is given by
\begin{equation}
\matr{M}_{\neu{}}=\left(\begin{array}{cccc} M_1 & 0 & -\MZ \sw \cos \beta& \MZ \sw \sin \beta\\ 0 & M_2 & \MZ \cw \cos \beta& -\MZ \cw \sin \beta\\ -\MZ \sw \cos \beta & \MZ \cw \cos \beta & 0 & -\mu\\\MZ \sw \sin \beta & -\MZ \cw \sin \beta & -\mu & 0\end{array}\right).
\label{neutralinomass1}
\end{equation}
The gluino is the only SUSY particle that enters only from
the two-loop level onwards; thus 
the impact of the gluino mass, $m_{\tilde{g}}=|M_3|$, on the $\MW$ prediction 
is relatively small.


\section{Determination of the \boldmath{$W$}~boson mass}
\label{sect:mwdet}

Muons decay via the weak interaction almost exclusively into 
$e \bar{\nu_{e}} \nu_{\mu}$~\cite{Beringer:1900zz}.  
The decay was originally described within the Fermi model, which
is a low-energy effective theory 
that emerges from the SM in the limit of vanishing
momentum transfer. The Fermi constant, $G_{\mu}$, 
is determined with high accuracy from precise measurements of the muon life 
time~\cite{Webber:2010zf}
and the corresponding Fermi-model prediction including QED corrections
up to $O(\alpha^2)$
for the point-like interaction~\cite{Behrends:1955mb,Kinoshita:1958ru,vanRitbergen:1999fi,Steinhauser:1999bx,Pak:2008qt}.
Comparison of the muon-decay amplitude in the Fermi model and in the SM or 
extensions of it yields the relation
\begin{equation}
\frac{G_{\mu}}{\sqrt{2}}=\frac{e^2}{8 s_W^2 \MW^2}\left(1+\Delta r\right) .
\label{eq:mwdef}
\end{equation}
Here $\Delta r$ represents the sum of all contributing loop diagrams to the 
muon-decay amplitude 
after splitting off the Fermi-model type virtual QED corrections,
\begin{equation}
\Dr = \sum_i \Dr_i\;,
\label{eq:drsum}
\end{equation}
with
\begin{equation}
{\cal M}_{\text{Loop,i}}=\Delta r_i \text{ }\mathcal{M}_{\text{Born}}\; .
\label{eq:drborn}
\end{equation}
This decomposition is possible
since after subtracting the Fermi-model QED corrections,
masses and momenta of the external fermions can be  neglected, 
which allows  
to reduce all loop contributions 
to a term proportional to the Born matrix element, 
see~\citeres{Sirlin:1980nh,Freitas:2002ja}.
By rearranging \refeq{eq:mwdef}, the $W$~boson mass can be calculated via 
\begin{equation}
\MW^2=\MZ^2 \left(\frac{1}{2}+\sqrt{\frac{1}{4}-\frac{\alpha \pi}
{\sqrt{2} G_{\mu} \MZ^2} \left(1 +\Delta r\right)}\right).
\label{eq:mwita}
\end{equation}
In different models, different particles can contribute as virtual
particles in the loop diagrams to the muon-decay amplitude. Therefore,
the quantity $\Delta r$ depends on the specific model parameters, and
\refeq{eq:mwita} provides a  
model-dependent prediction for the $W$~boson mass. 
The quantity $\Delta r$ itself does depend on $\MW$ as well; 
hence, the value of $\MW$ as the solution of \refeq{eq:mwita} has to
be determined numerically. In practice this is done by iteration. 
In most cases this procedure converges quickly 
and only a few iterations are needed. 

In order to exploit $\MW$ as a precision observable providing
sensitivity to quantum effects it is crucial that the theoretical
predictions for $\Dr$ are sufficiently precise with respect to the
present and expected future experimental accuracies of $\MW$.
Within the SM the full one-loop~\cite{Sirlin:1980nh,Marciano:1980pb} and 
two-loop~\cite{qcd2SMa,qcd2SMb,qcd2SMc,qcd2SMd,qcd2SMe,qcd2SMf,Freitas:2000gg, Freitas:2002ja, 2lfermc,2lbos,2Lbosa,2Lbosb}, 
as well as the leading higher-order corrections~\cite{Avdeev:1994db,qcd3SMa, qcd3SMb,Chetyrkin:1996cf,Faisst:2003px,Chetyrkin:2006bj,Boughezal:2006xk,vanderBij:2000cg,Boughezal:2004ef} are known.
In addition a convenient fitting formula for $\MW$ containing all numerically relevant contributions 
has been developed~\cite{Awramik:2003rn}, 
and in \citere{Awramik:2006uz} a corresponding formula for the
two-loop electroweak contributions to $\Delta r$ has been given.
In the MSSM the one-loop result
~\cite{Barbieri:1983wy,Lim:1983re,Eliasson:1984yu,Hioki:1985wz,
Grifols:1984xs,Barbieri:1989dc,Drees:1990dx,Drees:1991zk,
Chankowski:1994tn,Garcia:1993sb,Pierce:1996zz,Heinemeyer:2006px} 
and leading two-loop corrections have been 
obtained~\cite{Djouadi:1996pa,Djouadi:1998sq,Heinemeyer:2002jq,Haestier:2005ja}.


\section{Calculation of $\Dr$}\label{sect:deltar}

Our analysis is based on a new one-loop calculation of $\Dr$ in the 
MSSM with complex parameters which has been carried out using the {\tt
Mathematica}~\cite{mathematica} based programs
{\tt FeynArts} (Version 3.5)~\cite{Kublbeck:1990xc,Denner:1992me,Denner:1992vza,Kublbeck:1992mt,Hahn:2000kx,Hahn:2001rv}
and {\tt FormCalc} (Version 6.2)~\cite{Hahn:1998yk}, see
\citere{MWnmssm} for further details. 
The one-loop result is combined with all known higher order corrections
of SM and SUSY type as specified below, so that the numerical results
given in this paper correspond to the currently most precise predictions
for the $W$~boson mass in the SM and the MSSM.


\subsection{One-loop calculation in the MSSM}

The one-loop contributions to $\Delta r$ consist of the $W$~boson
self-energy, vertex and box diagrams, and the related counter terms (CT),
\begin{equation}
\begin{split}
 \Delta r\text{\hspace{0.1cm}}&= 
\text{\hspace{0.1cm}{$W$ Self-energy}\hspace{0.01cm}} 
+\text{\hspace{0.01cm}{$W$ Self-energy CT}\hspace{0.01cm}}
+\text{\hspace{0.01cm}Vertex\hspace{0.01cm}}
+\text{\hspace{0.01cm}{Vertex CT}\hspace{0.01cm}}
+\text{\hspace{0.01cm}Box\hspace{0.1cm}}\\
 &={\frac{\Sigma_T^{WW}\left(0\right)}{\MW^2}}
+{\left(-\delta Z_W-\frac{\delta \MW^2}{\MW^2}\right)}
+\text{\hspace{0.05cm}Vertex\hspace{0.05cm}}\\
&+{\left(2 \delta Z_e-2\frac{\delta \sw}{\sw}
+\delta Z_W+\frac{1}{2}\left(\delta Z^{\mu}+\delta Z^{e}
+\delta Z^{\nu_{\mu}}+\delta Z^{\nu_{e}}\right)\right)}
+\text{\hspace{0.05cm}Box\hspace{0.05cm}}.
\label{herleitungdeltar}
\end{split}
\end{equation}
Here $\Sigma_T$ denotes the transverse part of a gauge boson
self-energy, $\delta \MW$ is the counter\-term for the $W$ boson mass, 
$\delta Z_e$ and $\delta \sw$ are the renormalization constants for the
electric charge and the (sine of the) weak mixing angle, respectively,
while the other
$\delta Z$ denote field renormalization constants.
Since the $W$~boson appears only as a virtual particle, its field
renormalization constant $\delta Z_W$ drops out in the $\Delta r$ formula. 
The box diagrams are themselves UV-finite in a renormalizable gauge.
Choosing on-shell renormalization conditions,%
\footnote{The on-shell renormalization conditions correspond to the
definition of the $W$ and $Z$ boson masses according to the real part
of the complex pole of the propagator. This gives rise to the fact that
the predictions for $\Dr$ discussed in this paper internally make use of
a definition of the gauge boson masses in terms of a Breit--Wigner
shape with a fixed width. The values of the $W$ and $Z$ boson masses
according to this fixed-width definition are finally converted into the
running-width definition which has been adopted for the determination of
the experimental values of $\MW$ and $\MZ$, see e.g.\
\citere{Freitas:2002ja} for further details.}
which ensures that 
\refeq{eq:mwdef} corresponds to the relation between the physical masses
of the $W$ and $Z$~bosons, yields (neglecting the masses of the external
fermions)%
\footnote{We adopt here the sign conventions for the covariant
derivative used in {\tt FeynArts}~\cite{Kublbeck:1990xc,Denner:1992me,Denner:1992vza,Kublbeck:1992mt,Hahn:2000kx,Hahn:2001rv}, which are different for the SM and
the MSSM. Accordingly,
$\text{sgn}$ (the sign of the term involving the SU(2) coupling in the
covariant derivative) in
\refeq{eq:deltar1LOS} for this choice of convention is
$\text{sgn}=-1$ in the SM and $\text{sgn}=+1$ in the MSSM.
\refeq{eq:deltar1LOS} agrees with the corresponding formula given in
\citere{Heinemeyer:2006px} up to typographical errors in
\citere{Heinemeyer:2006px}.}%
\begin{equation}
\begin{split}
\Delta r = &\frac{\Sigma_T^{WW}(0)-\text{Re} \left(\Sigma_T^{WW}(\MW^2)\right)}
                 {\MW^2}+\Pi^{AA}\left(0\right)
-\frac{\cw^2}{\sw^2}\,  \text{Re} 
\left[\frac{\Sigma^{ZZ}_T(\MZ^2)}{\MZ^2}
     -\frac{\Sigma^{WW}_T(\MW^2)}{\MW^2}\right]\\
&+ 2 \;\frac{ \text{sgn} \; \sw}{\cw} \frac{\Sigma^{AZ}_T(0)}{\MZ^2} 
+\text{Vertex}+\text{Box}-\frac{1}{2} \text{Re} 
\left( \Sigma^{e}_L(0) +\Sigma^{\mu}_L(0)
+\Sigma^{\nu_{e}}_L(0) +\Sigma^{\nu_{\mu}}_L(0) \right), 
\end{split}    
\label{eq:deltar1LOS}
\end{equation} 
with the photon vacuum polarization
\begin{equation}
\Pi^{AA}(k^2)=\frac{\Sigma_{T}^{AA}(k^{2})}{k^2}.
 \label{eq:pi1}
\end{equation}
Here $\Sigma_L$ denotes the left-handed part of a fermion
self-energy.

The contributions to $\Dr$ in the MSSM, besides the ones of SM type, 
consist of a large number of additional self-energy, vertex and box diagrams 
containing
sfermions, (SUSY) Higgs bosons, charginos and neutralinos in the loop,
see also \citere{Heinemeyer:2006px}.
In order to determine
the contribution to $\Delta r$ from a particular loop diagram, 
the Born amplitude has to be factored out of the one-loop muon decay amplitude, as shown in~\refeq{eq:drborn}. 
While most loop diagrams directly give a result proportional to the Born
amplitude, more complicated spinor structures that do not occur in
the SM case arise from box diagrams containing neutralinos and
charginos. Those spinor chains can be related to the Born amplitude with
the help of Fierz identities and charge conjugation relations. The
reduction of the box diagrams to Born-type amplitudes leads to
coefficients containing ratios of mass-squared differences of the
involved particles. These coefficients can give rise to numerical
instabilities in cases of mass degeneracies. In the implementation of
our results (which has been carried out in a {\tt Mathematica} and a {\tt
Fortran} version) special care has been taken of such parameter regions
with mass degeneracies or possible threshold effects, so that a 
numerically stable evaluation is ensured.

At the one-loop level, the quantity $\Delta r$ can be split into three parts
\begin{equation}
\Delta r^{(\al)}=\Delta \alpha-\frac{\cw^2}{\sw^2} \Delta \rho + 
\Delta r_{\text{rem}}.
\label{eq:deltarSM1L2}
\end{equation} 
The shift of the fine structure constant $\Delta \alpha$
arises from the charge renormalization
which contains the contributions from light fermions.
The quantity
$\Delta \rho$ contains loop corrections to the $\rho$
parameter~\cite{Veltman:1977kh}, 
which describes the ratio between neutral and charged weak currents, and can be written as
\begin{equation}
\Delta \rho = \frac{\Sigma^{ZZ}_T (0)}{M^2_Z}- 
\frac{\Sigma^{WW}_T (0)}{M^2_W}\;.
 \label{eq:deltarho}
\end{equation} 
This quantity is sensitive to the mass splitting between the isospin partners in a doublet~\cite{Veltman:1977kh},
which leads to a sizable effect in the SM
in particular from the heavy fermion doublet.
While $\Delta \alpha$ is a pure SM contribution, $\Delta \rho$ can
get large contributions also from SUSY particles, in particular 
the superpartners of the heavy quarks.
All other terms, both of SM and SUSY type, are contained in the remainder term $\Delta r_{\text{rem}}$.


\subsection{Incorporation of higher order corrections}
\label{sect:hocorrections}

The one-loop result described above has been combined 
with all available higher-order corrections. 
Since the calculation of 
$\De r$ in the SM is more advanced than in the MSSM 
we have organized our result such that the full SM result for $\De r$
can be used also for the MSSM prediction of $\MW$.
Therefore the MSSM result is split into a SM part and a SUSY part%
\footnote{Since the complete one-loop results for $\De r$ in the SM and
in the MSSM are used in \refeq{eq:deltarsplit}, this splitting has an
impact only from the two-loop level onwards.}
\begin{equation}
\begin{split}
&\Delta r^{\text{MSSM}}=\Delta r^{\text{SM}}+\Delta r^{\text{SUSY}}\;.
\end{split}    
\label{eq:deltarsplit}
\end{equation}
Writing the MSSM result in terms of \refeq{eq:deltarsplit} ensures
in particular that in the decoupling limit of the MSSM result, where all
superpartners are heavy and the Higgs sector becomes SM-like, the 
full SM result (with $\MHSM = \Mh$) is recovered, see also the
discussion in~\citere{Heinemeyer:2006px}.
The SM part of $\Dr$ up to four-loop order is given by
\begin{equation}
\begin{split}
 \Delta r^{\text{SM}}=&\Delta r^{(\alpha)}+\Delta r^{(\alpha \alpha_s)}
+\Delta r^{(\alpha \alpha_s^2)}+\Delta r_{\text{ferm}}^{(\alpha^2)}+\Delta r_{\text{bos}}^{(\alpha^2)}\\
 &+\Delta r^{(G_{\mu}^2 \alpha_s \mt^4)}+\Delta r^{(G_{\mu}^3 \mt^6)}+\Delta r^{(G_{\mu} \mt^2 \alpha_s^3)}\;.
\end{split}
 \label{eq:SMhighercont}
\end{equation} 
It contains, besides the one-loop contribution $\Delta r^{(\alpha)}$, 
\begin{itemize}
\item the two-loop QCD corrections $\Delta r^{\alpha \alpha_s}$ 
\cite{qcd2SMa,qcd2SMb,qcd2SMc,qcd2SMd,qcd2SMe,qcd2SMf},
\item the three-loop QCD corrections $\Delta r^{\alpha \alpha_s^2}$ 
\cite{Avdeev:1994db,qcd3SMa,qcd3SMb,Chetyrkin:1996cf},
\item the fermionic electroweak two-loop corrections $\Delta r_{\text{ferm}}^{(\alpha^2)}$ \cite{Freitas:2000gg,Freitas:2002ja,2lfermc},
\item the purely bosonic electroweak two-loop corrections $\Delta r_{\text{bos}}^{(\alpha^2)}$ \cite{2lbos,2Lbosa,2Lbosb},
\item the mixed QCD and electroweak three-loop contributions $\Delta r^{G_{\mu}^2 \alpha_s \mt^4}$ \cite{Faisst:2003px,vanderBij:2000cg}, 
\item the purely electroweak three-loop contribution $\Delta r^{G_{\mu}^3 \mt^6}$ \cite{Faisst:2003px,vanderBij:2000cg}, 
\item and the four-loop QCD correction $\Delta r^{(G_{\mu} \mt^2 \alpha_s^3)}$ \cite{Boughezal:2006xk}. 
\end{itemize}
The full result for the 
electroweak two-loop contributions in the SM involves numerical
integrations of the two-loop scalar integrals, which make the
corresponding code rather unwieldy and slow. Thus, we make use of the 
simple parametrisation that has been given in~\citere{Awramik:2006uz}
for the combined result of the fermionic and bosonic electroweak
two-loop corrections in the SM, which approximates
the exact result for $\Delta r_{\text{ferm}}^{(\alpha^2)} + \Delta
r_{\text{bos}}^{(\alpha^2)}$ to better than $2.7 \times 10^{-5}$ for
$10 \gev \leq \MHSM \leq 1 \tev$ (and the other input parameters in
their $2\,\sigma$ ranges), corresponding to an uncertainty of 
$0.4 \mev$ for $\MW$. 
The use of a parametrisation directly for the SM prediction of 
$\Delta r_{\text{ferm}}^{(\alpha^2)} + \Delta r_{\text{bos}}^{(\alpha^2)}$ 
rather than for the full SM prediction of $\MW$ leads to an improved
accuracy in the combination with the SUSY contributions as compared to
\citere{Heinemeyer:2006px}.
Concerning the QCD corrections, which enter from the two-loop level
onwards, it should be noted that they result in a rather 
large (downward) shift of the $W$ boson mass prediction. It is obvious
that this kind of corrections needs to be theoretically well under
control in order to gain sensitivity to effects of physics beyond the SM.

The quantity
$\Delta r^{\text{SUSY}}$ in \refeq{eq:deltarsplit} denotes 
the difference between $\Delta r$ in the
MSSM and the SM, i.e.\ it only involves the contributions from the
additional SUSY particles and the extended Higgs sector. Beyond one-loop
order, all SUSY corrections that are known to date are implemented,
namely the leading reducible ${\cal O}(\alpha^2)$ two-loop corrections
that can be obtained via the resummation formula given in
\citere{Consoli:1989fg}, the leading SUSY two-loop QCD
corrections of ${\cal O}(\alpha \alpha_s)$ to $\Delta \rho$ as given in
\citeres{Djouadi:1996pa,Djouadi:1998sq}, as well as
the dominant Yukawa-enhanced electroweak corrections of
${\cal O}(\alpha_t^2)$, ${\cal O}(\alpha_t \alpha_b)$, ${\cal O}(\alpha_b^2)$ 
to $\Delta \rho$~\cite{Heinemeyer:2002jq,Haestier:2005ja}.
In order to incorporate the latter corrections, the dominant
Yukawa-enhanced electroweak corrections in the
SM~\cite{Barbieri:1992dq,Fleischer:1993ub} have been subtracted from the
MSSM result presented in \citere{Haestier:2005ja} according to
\refeq{eq:deltarsplit}. For this purpose we have identified the SM Higgs
mass entering the result of \citeres{Barbieri:1992dq,Fleischer:1993ub}
with the mass of the MSSM Higgs boson that has the largest coupling to
gauge bosons (i.e., the MSSM Higgs boson that behaves most SM-like).
In the decoupling limit, where $\MA \gg \MZ$ and all superpartners are
heavy, the MSSM contribution reduces to the SM contribution with 
$\MHSM = \Mh$, so that the contribution to $\Delta r^{\text{SUSY}}$ 
vanishes as required.


\section{Numerical analysis}
\label{sect:numerics}

Our numerical results are based on the contributions to
$\Dr$ described in the previous section (which have been implemented in a 
{\tt Mathematica} and a {\tt Fortran} version, where the latter has been used
to generate the results presented below). The numerical values for the
masses and effective couplings of the MSSM Higgs bosons have been
evaluated with the help of the program {\tt FeynHiggs} (version
2.9.4)~\cite{Hahn:2009zz,Frank:2006yh,Degrassi:2002fi,
Heinemeyer:1998np,Heinemeyer:1998yj}.
We cross-checked our evaluation with the earlier results given
in~\citere{Heinemeyer:2006px} and found good agreement, at the level of
about $1$--$2\mev$.

\subsection{Prediction for \boldmath{$\MW$} in the SM}

The mass of the signal discovered in the Higgs boson searches at the LHC
about a year ago is measured mainly in the $\ga \ga$ and the $ZZ^{(*)}$
channels. Currently, the combined mass measurement from ATLAS is $125.5
\pm 0.2 \pm 0.6 \GeV$~\cite{ATLAS:2013mma}
 and from CMS $125.7 \pm 0.3 \pm 0.3 \GeV$ \cite{CMS:yva}. 
Adding systematic and statistical errors in quadrature and
determining
the weighted average between the ATLAS and CMS measurements we get
$\MHSM = 125.64 \pm 0.35 \gev$.
Setting the SM Higgs boson mass to this value, the SM prediction for the 
$W$ boson mass reads
(the other SM parameters have been fixed as
$G_{\mu} = 1.1663787 \times 10^{-5}$, 
$\MZ =91.1875 \gev$, 
$\als(\MZ) =0.1180$, 
$\De\al_{\rm had} =0.02757$)
\begin{equation}
\MW^{\rm SM}(\mt= 173.2 \GeV, \MHSM = 125.64 \GeV) = 80.361 \GeV.
\label{eq:mwsmresult}
\end{equation} 
Accordingly, the SM prediction for $\MW$ turns out to be below the
current experimental value, $\MW^{\rm exp} = 80.385 \pm 0.015 \gev$,
by about $1.5\,\sigma$.
The dominant theoretical uncertainty of the prediction for $\MW$ 
arises from the parametric uncertainty induced by
the experimental error in the measurement of the top-quark mass.
An experimental error of 1 GeV on $\mt$ causes a parametric 
uncertainty on $\MW$ of about $6 \mev$, while the parametric
uncertainties induced by the current
experimental error of the hadronic contribution to the shift in the
fine-structure constant, $\De\al_{\rm had}$, and by the experimental
error of $\MZ$ amount to about $2\mev$ and $2.5\mev$, respectively.
The uncertainty of the $\MW$ prediction caused by the experimental 
error of the Higgs mass $\de\MH^{\rm exp} = 0.35 \gev$ is significantly smaller ($\sim 0.2 \mev$).
The uncertainties from unknown higher-order corrections have
been estimated to be around $4$~MeV in the SM for a light Higgs
boson ($\MHSM < 300 \gev$)~\cite{Awramik:2003rn}.  


\subsection{MSSM parameter scan: Scan ranges and constraints}
\label{sec:constraints}

The prediction for $\MW$ in the MSSM is affected by additional
theoretical uncertainties from unknown higher-order corrections of SUSY
type. While in the decoupling limit those additional uncertainties
vanish, they can be important if some SUSY particles, in particular in
the scalar top and bottom sectors, are relatively light. The combined
theoretical uncertainty from unknown higher-order corrections of SM- and
SUSY-type has been estimated (for the MSSM with real parameters) in 
\citeres{Haestier:2005ja,Heinemeyer:2006px} as
$\delta \MW = (4.7-9.4)$~MeV, depending on the SUSY mass scale.

In the following we will investigate the prediction for $\MW$ in the
MSSM based on scans of the MSSM parameters over a wide range (using flat
distributions).  We have performed two versions of those random scans, 
one where the top-quark mass is kept fixed at $\mt = 173.2 \gev$ and one
where also $\mt$ is allowed to vary in the scan.
Both scans use initially $\sim 5\times 10^6$ points, and dedicated
smaller scans have been performed in parameter regions where the SUSY
contributions to $\MW$ are relatively large.
The scan ranges are given in \refta{tab:scanparam}. 
We have assumed that the value of $M_1$ is fixed by the one of
$M_2$ in terms of the usual GUT relation,
$M_1 = 5/3\, \sw^2/\cw^2\, M_2$.
As mentioned above, we restrict our numerical analysis to the 
case of real parameters.
We include CKM mixing, but the numerical effect turns out to
be negligible (below $0.01 \mev$ in $\MW$). 
Possible flavor violation in the SUSY
sector~\cite{Heinemeyer:2004by} is neglected here. 
In order to avoid unphysical parameter regions and regions of
numerical instabilities we disregard parameter points for which 
{\tt FeynHiggs} indicates a large theoretical uncertainty in the
evaluation of the Higgs mass predictions. We furthermore 
exclude points where stop and sbottom masses are mass-degenerate within
less than $0.1\gev$ causing numerical instabilities in the gluino
corrections of ${\cal O}(\alpha \alpha_s)$ to $\Delta \rho$.

\begin{table}[htb!]
\centering
\begin{tabular}{cccl}
\hline
Parameter &  Minimum &  Maximum \\
\hline
$\mu$ & -2000       & 2000 \\
$M_{\tilde{E}_{1,2,3}}=M_{\tilde{L}_{1,2,3}}$ & 100       & 2000 \\
$M_{\tilde{Q}_{1,2}}=M_{\tilde{U}_{1,2}}=M_{\tilde{D}_{1,2}}$ & 500       & 2000 \\
$M_{\tilde{Q}_{3}}$     & 100       & 2000 \\
$M_{\tilde{U}_{3}}$     & 100       & 2000 \\
$M_{\tilde{D}_{3}}$     & 100       & 2000 \\
$A_e=A_{\mu}=A_{\tau}$   & -3$\,M_{\tilde{E}}$      & 3$\,M_{\tilde{E}}$     \\
$A_{u}=A_{d}=A_{c}=A_{s}$& -3$\,M_{\tilde{Q}_{12}}$  & 3$\,M_{\tilde{Q}_{12}}$ \\
$A_b$ & -3$\,$max($M_{\tilde{Q}_{3}},M_{\tilde{D}_{3}}$)  &
3$\,$max($M_{\tilde{Q}_{3}},M_{\tilde{D}_{3}}$) \\ 
$A_t$ & -3$\,$max($M_{\tilde{Q}_{3}},M_{\tilde{U}_{3}}$)  &
3$\,$max($M_{\tilde{Q}_{3}},M_{\tilde{U}_{3}}$) \\ 
$\tb$     & 1       & 60 \\
$M_3$     & 500     & 2000 \\
$M_A$     & 90      & 1000\\
$M_2$     &100      &1000\\
\hline
\end{tabular}
\caption{Parameter ranges considered in the scans. 
All parameters with mass dimension are given
  in GeV.
}
\label{tab:scanparam}
\end{table}

All MSSM points included in our results have the lightest
neutralino as LSP
and have SUSY particle masses that 
pass the lower mass limits from direct searches
at LEP.  
The Higgs and SUSY masses are calculated from the MSSM input
parameters using {\tt FeynHiggs}
(version 2.9.4)~\cite{Frank:2006yh,Degrassi:2002fi,Heinemeyer:1998np,Heinemeyer:1998yj}. 
In the SM and SUSY higher-order corrections, as listed in
\refse{sect:hocorrections}, the bottom-quark mass has been renormalized
in the on-shell scheme. Accordingly, in our evaluation of $\MW$ the
bottom-quark pole mass, $m_b^{\text{pole}}$, is used everywhere. This
also applies to the calculation of the 
sbottom masses from the MSSM input parameters, and we have modified the
corresponding routine in {\tt FeynHiggs} accordingly (in the calculation
of the sbottom masses furthermore a 
$\Delta_b$~\cite{Hempfling:1993kv,Hall:1993gn,Carena:1994bv,Carena:1999py}
correction enters, which can be absorbed into an effective bottom-quark
mass).
For every parameter point we test whether it is allowed by direct 
Higgs searches using the code {\tt HiggsBounds}  
(version 4.0.0)~\cite{Bechtle:2013gu,Bechtle:2011sb,Bechtle:2008jh}. 
This code tests the compatibility of the MSSM points with the
search
limits from LEP, the Tevatron and the LHC.
Running {\tt HiggsBounds}, we take into account the theoretical 
uncertainties on the Higgs masses using the estimate provided by
{\tt FeynHiggs}.

Our results presented below improve on earlier results given in 
\citere{Heinemeyer:2006px} in several respects. We study here the impact
of both the limits from the Higgs boson searches as well as from the
signal observed at about $125.6 \gev$. Furthermore we investigate
constraints from present and possible future limits from searches for
SUSY particles. On a more technical level, our analysis incorporates the 
SUSY two-loop corrections of ${\cal O}(\alpha_t^2)$, 
${\cal O}(\alpha_t \alpha_b)$, ${\cal O}(\alpha_b^2)$, which were not 
included in the scan results presented previously, and we perform a more
detailed scan involving a larger number of sampling points.


\subsection{Results for \boldmath{$\MW$} in the MSSM}
\label{sec:numanal}
\begin{figure}
\centering
\begin{tikzpicture}
    \node[anchor=south west,inner sep=0] at (0,0) {
    \includegraphics[width=0.48\columnwidth]{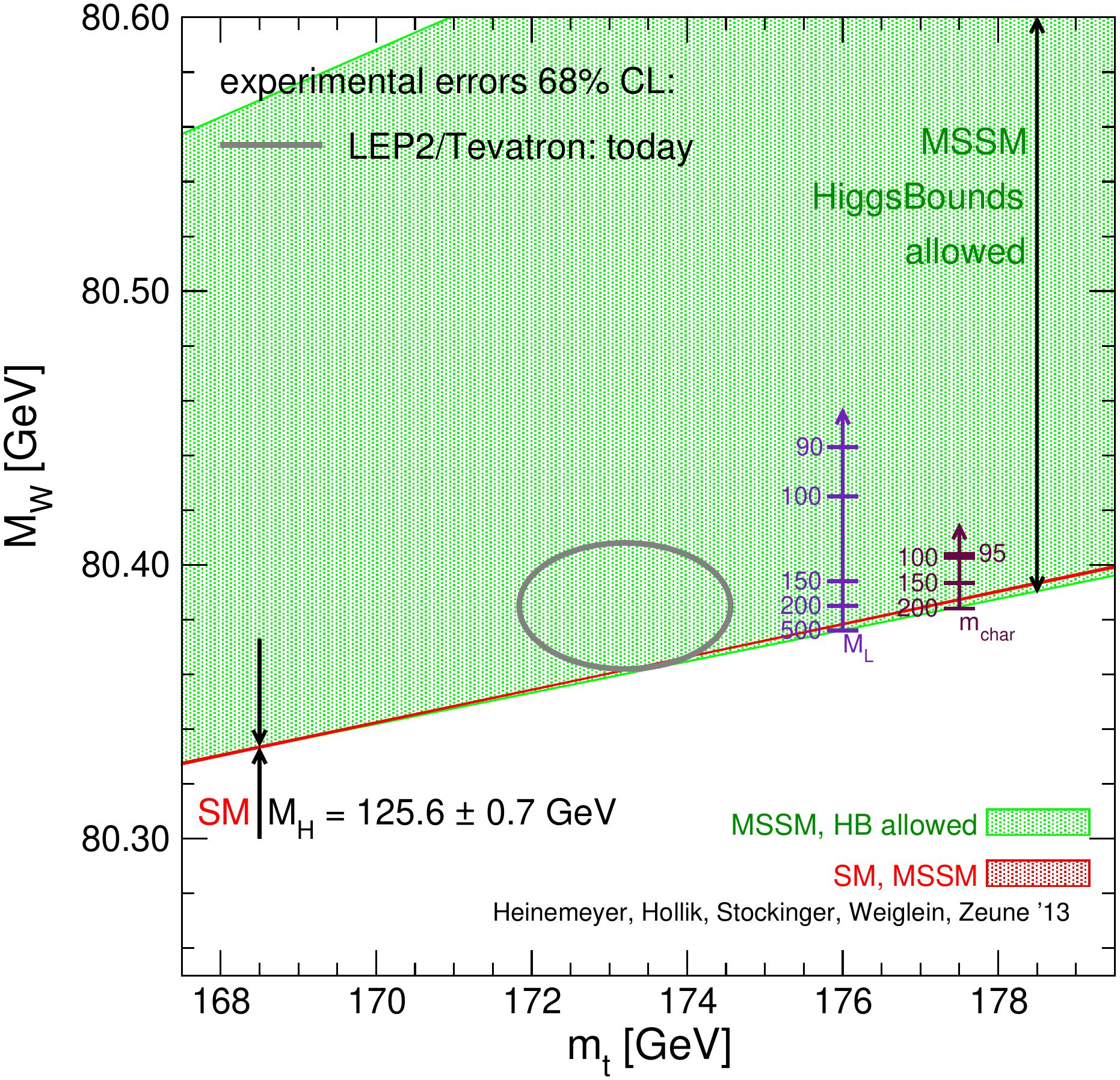}
    \includegraphics[width=0.49\columnwidth]{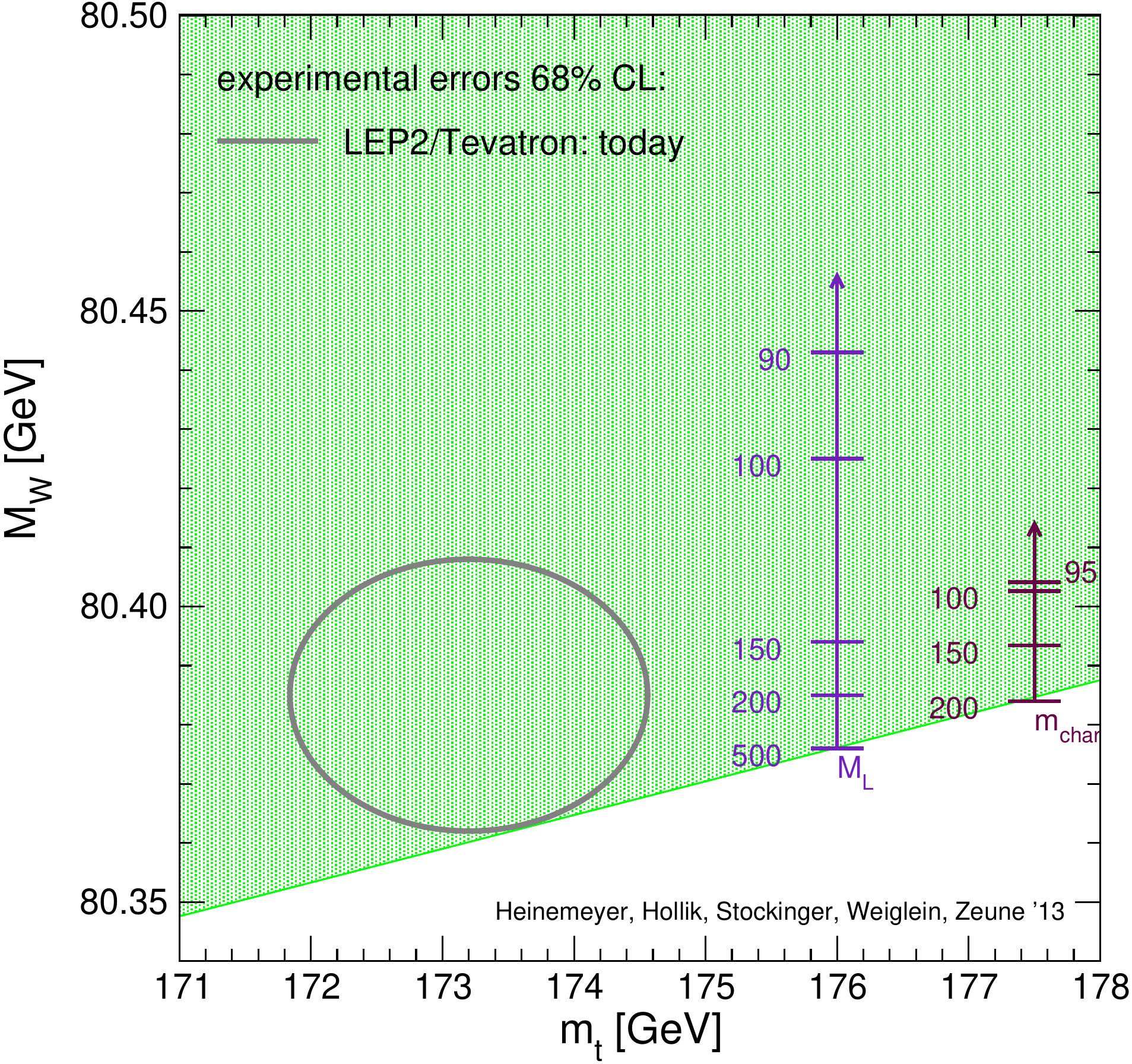}};
    \draw[fill=white,draw=white] (3.2,1) rectangle (7.7,1.3);
    \draw[fill=white,draw=white] (11.3,0.95) rectangle (15.7,1.3);
\end{tikzpicture}
\caption{Prediction for $\MW$ as a function of $\mt$. 
 Left: The green region shows the \HiggsBounds\ allowed region for the MSSM $\MW$ prediction. It has been obtained by scanning over the MSSM parameters as described in the text. 
The cuts $m_{\tilde{t}_2}/m_{\tilde{t}_1}<2.5$ and $m_{\tilde{b}_2}/m_{\tilde{b}_1}<2.5$ are applied.
The red strip indicates the overlap region of the SM and the MSSM, with
$\MHSM = 125.6 \pm 0.7 \gev$. 
 The two arrows indicate the possible size of the
  slepton and the chargino (and neutralino) contributions. 
Right: zoom into the most relevant region, with the SM area omitted.
}
\label{fig:mtmwFull}
\end{figure}

In this section we study the MSSM prediction for $\MW$, starting in 
\reffi{fig:mtmwFull} where $\MW$ is displayed as a function of the 
top-quark mass, $\mt$, in the SM and the MSSM. The green area shows the
MSSM parameter space that is allowed by \HiggsBounds\ and the various other
constraints described in the previous subsection. It should be noted
that in this plot only the limits from the Higgs searches are considered
as constraints on the MSSM parameter space, not the observed signal at
about $125.6 \gev$ (the latter will be discussed below). The region where the
MSSM prediction for $\MW$ overlaps with the one in the SM is indicated
by the red strip, where $\MHSM = 125.6 \pm 0.7 \gev$
(corresponding roughly to the $2\,\sigma$ experimental error on $\MH$)
has been used for the SM prediction. The left plot shows the results on
a larger scale, in order to indicate the possible range of the MSSM
prediction, while the right plot is a zoom into the parameter region of
the MSSM near the experimental central values of $\MW$ and $\mt$.
In order to obtain the MSSM prediction shown as the green band in
\reffi{fig:mtmwFull} we have imposed as an additional restriction a
limit on the mass splittings in the stop and sbottom sector, which has
been implemented via the conditions
$m_{\tilde{t}_2}/m_{\tilde{t}_1}<2.5$ and 
$m_{\tilde{b}_2}/m_{\tilde{b}_1}<2.5$. If no such condition on the mass
splittings in the stop and sbottom sector were imposed, even larger
values of $\MW$ (up to $\sim 80.8 \gev$) would be possible in the MSSM,
see also the discussion in \citere{Heinemeyer:2006px}. Since this
parameter region far above the experimental value of $\MW$ is of little
phenomenological interest, we will not consider it further here.
While it is well-known that a non-zero SUSY contribution tends to
increase the prediction for $\MW$ as compared to the SM case, close
inspection of \reffi{fig:mtmwFull} reveals that there exists a small
MSSM (green) region below the overlap region between the MSSM and the 
SM (red), which is best visible for the largest $\mt$ values. 
The reason for this feature lies in the fact that, as explained above,
the SM prediction is shown for the range $\MHSM = 125.6 \pm 0.7 \gev$,
while no restriction from the signal observed in the Higgs searches has
been applied to the MSSM parameter space. As a consequence, the MSSM
region (green) contains parameter points where the lightest $\cp$-even
Higgs boson of the MSSM has a mass above the range allowed for $\MHSM$ 
(and below the upper bound on $\Mh$ in the MSSM, which increases with
increasing $\mt$). In the decoupling region, where all superpartners are
heavy, the MSSM prediction for $\MW$ in this case corresponds to the
prediction in the SM with a higher value of $\MHSM$, which yields a lower
value of $\MW$%
\footnote{It should be noted that a similar kind of feature would occur
even if one restricted the predicted value for $\Mh$ in the MSSM to the
same region as the range adopted for $\MHSM$. This is caused by the fact
that the additional theoretical uncertainties from unknown higher-order
corrections affecting the prediction for $\Mh$ in the MSSM, which are
not present in the SM where $\MHSM$ is a free input parameter,
essentially lead to a broadening of the allowed range of $\Mh$ in the
MSSM as compared to $\MHSM$.}.

The predictions for $\MW$ in the SM and the MSSM are compared with
the current experimental results for $\MW$ and
$\mt$~\cite{Group:2012gb} which are displayed by the
corresponding 68\%~C.L.\ ellipse shown in gray.
One can see that the SM prediction barely touches the 68\%~C.L.\ ellipse, 
whereas the ellipse is fully contained in the MSSM area. 
It is obvious that the MSSM contains parameter regions where the MSSM
prediction for $\MW$ is in very good agreement with the data. On the
other hand, also $\MW$ values significantly above the experimental value
are possible in the MSSM. The latter arise mainly from very light states
and a large mass splitting in the stop and sbottom sector (see the
discussion below). 

\reffi{fig:mtmwFull} shows that confronting the prediction for $\MW$ in
the MSSM with the experimental result is of interest both for 
putting constraints on parameter regions that would give rise to a too
high value of $\MW$ and for 
investigating the parameter region where the agreement between the MSSM
prediction and the data is in fact better than for the SM case.
While the deviation between the SM prediction and the experimental
result for $\MW$ is statistically not very significant (the SM
prediction is well compatible with the experimental result at the
95\%~C.L.), the pattern that the SM prediction is somewhat low as
compared to the data has been robust for many years in spite of numerous
updates of the experimental results.
Focussing now on the region where we find the best
agreement between the MSSM prediction for $\MW$ and the experimental
result, it is interesting to note that in this region some of the
superpartner masses are expected to be relatively light. In order to
illustrate this feature we furthermore show in \reffi{fig:mtmwFull}
the impact of the slepton sector (left arrow) and the chargino sector 
(right arrow), where the mass values indicated at the arrows
(approximately) show the effect in $\MW$ arising from the contribution 
of a slepton and a chargino having this mass, respectively.
We have chosen to display those arrows such that they start at the lower
border, corresponding to the situation where all other superpartners are
heavy and decoupled.
For the sleptons we show the
corrections to $\MW$ as a function of 
$M_L \equiv M_{\tilde{E}_{1,2,3}}=M_{\tilde{L}_{1,2,3}}$, where the
lower limit of $\sim 90 \gev$ roughly corresponds to the (fairly
model-independent) limit
obtained at LEP. One can see that very light sleptons, just
above the LEP limit, could induce a shift in $\MW$ of about $60 \mev$.
We have checked that each generation contributes
roughly the same to this effect.
The major contributions to $\MW$ 
from the sleptons arise from the $\Delta \rho$ term in 
\refeq{eq:deltarSM1L2}, which is sensitive to the mass splitting between 
$\tilde{l}_{1,2}$ and $\tilde{\nu_l}$. The splitting between the sneutrinos and
the sleptons becomes significant if $M_{\tilde{E}}=M_{\tilde{L}}$ and
$\MW$ are of comparable size.
The contributions to $\MW$ from light
charginos and neutralinos are substantially smaller, but clearly not negligible 
in this context. They reach about 
$20 \mev$ for $m_{\cha{1}} \sim 95 \gev$,
close to its lower mass limit from LEP.  
In that case, due to the assumed 
GUT relation between $M_1$ and $M_2$,
the mass of $\neu{1}$ is $\sim 50 \gev$.
Our analysis of the contributions in the slepton and the chargino /
neutralino sector shows that even if all squarks were so heavy that
their contribution to the $\MW$ prediction were negligible,
contributions from the slepton sector or the chargino / neutralino
sector could nevertheless be sufficient to bring the MSSM prediction in
perfect agreement with the data. This could be the case for slepton
masses of about $150$--$200\gev$ or for a chargino mass of
about $100$--$150\gev$. If the squark sector gives rise to a non-zero
contribution to $\MW$ the same predicted value for $\MW$ could be
reached with heavier sleptons and charginos / neutralinos.

\begin{figure}
\centering
\includegraphics[width=0.48\columnwidth]{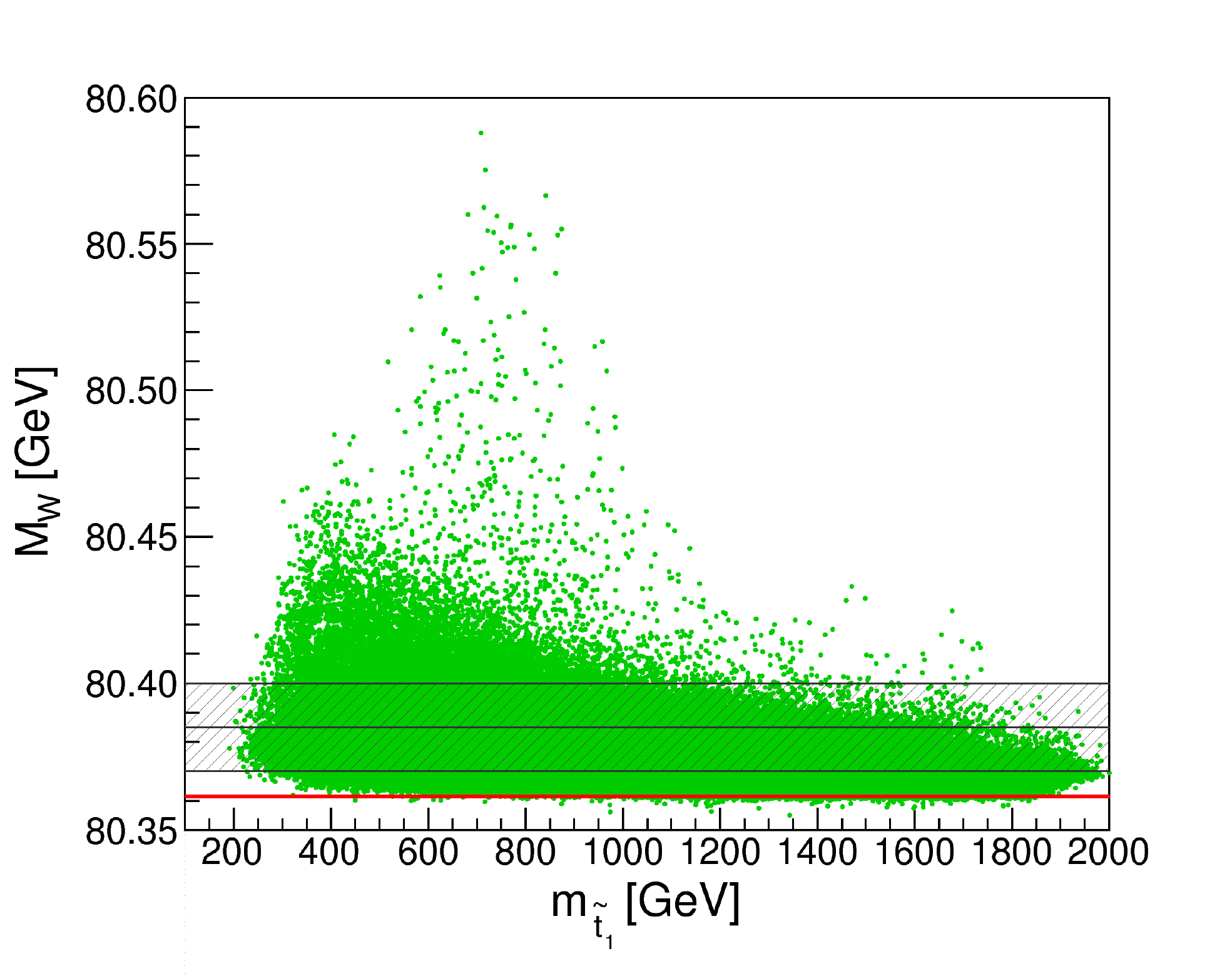}
\includegraphics[width=0.48\columnwidth]{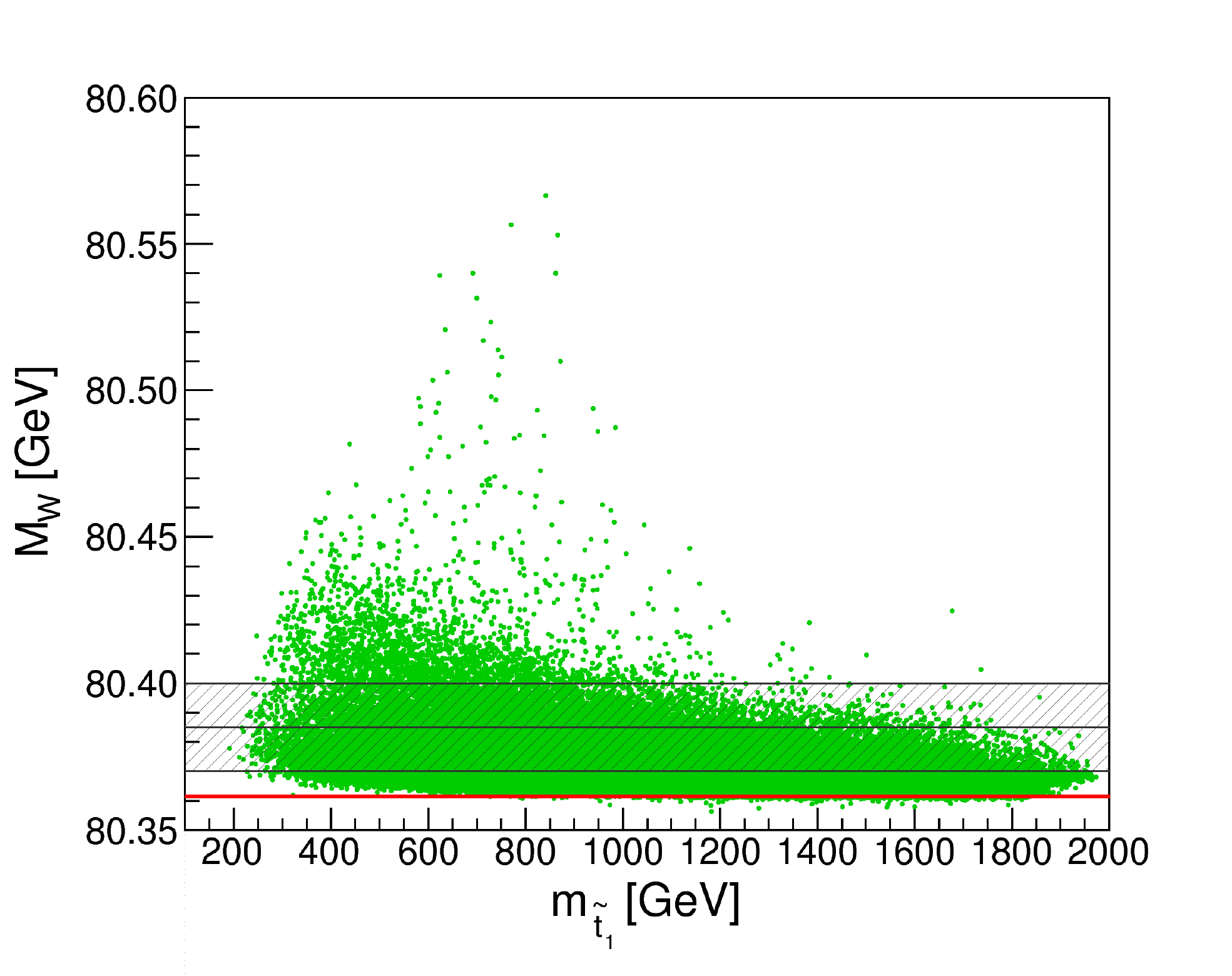}
\includegraphics[width=0.48\columnwidth]{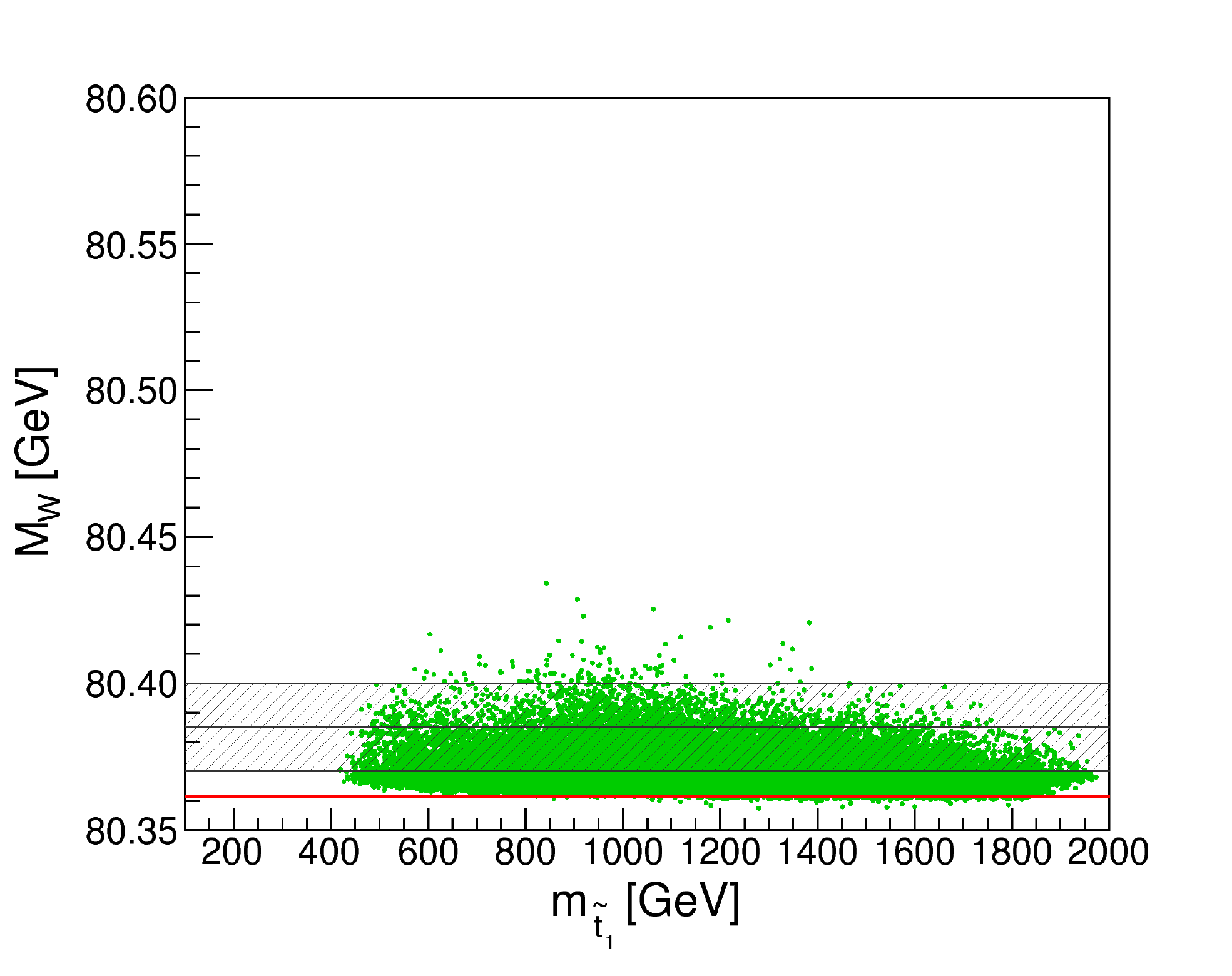}
\includegraphics[width=0.48\columnwidth]{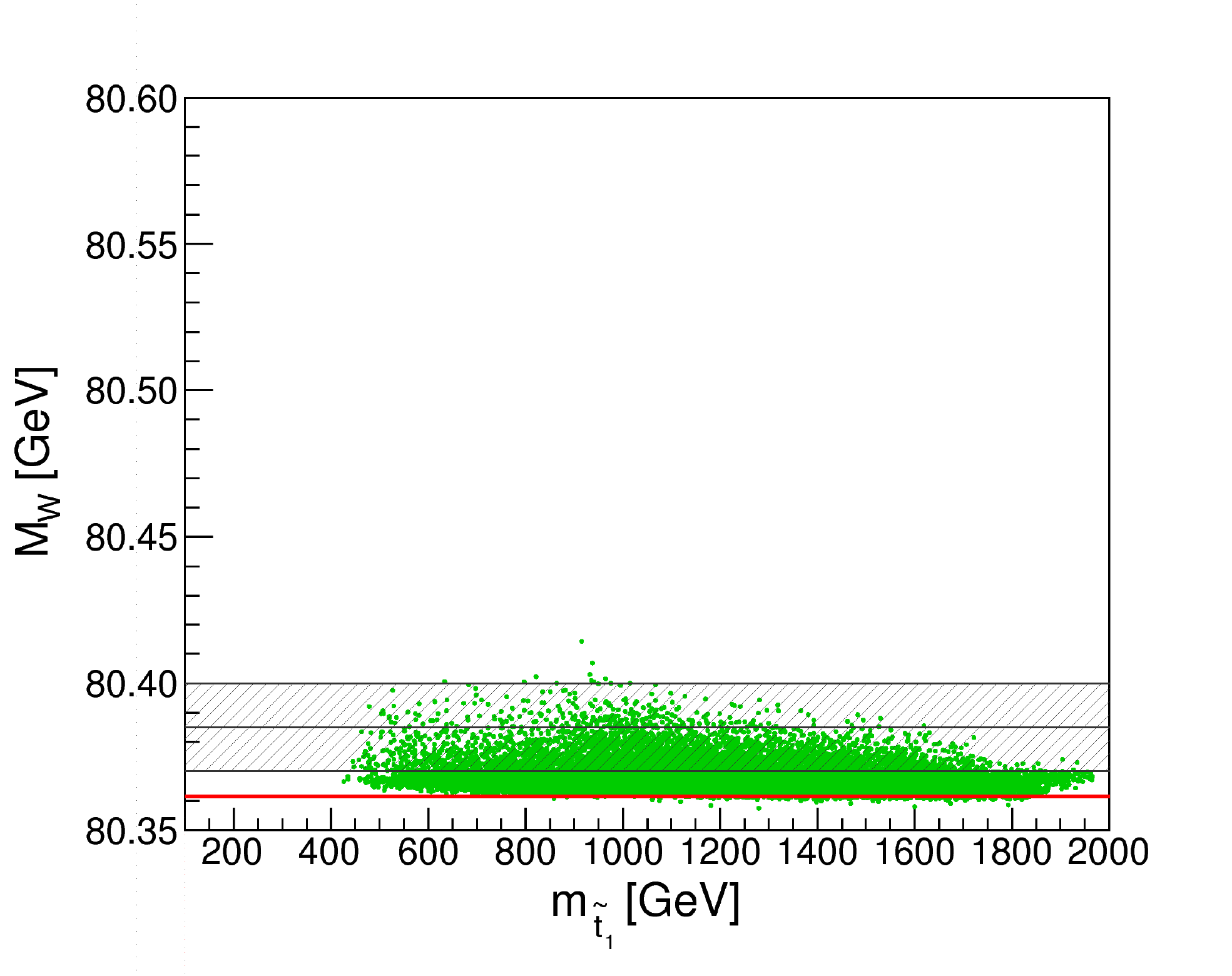}
\caption{Prediction for $\MW$ as a function of the lightest stop mass $\mste$.  
In all plots the cuts $m_{\tilde{t}_2}/m_{\tilde{t}_1}<2.5$ and $m_{\tilde{b}_2}/m_{\tilde{b}_1}<2.5$ are applied.
In the upper left plot all \HiggsBounds\ allowed points are shown, 
in the upper right plot only the points are shown for which additionally
the squarks of the first two generations and the gluino are heavier
than $1200 \gev$,  
in the lower left plot only the points are shown for which additionally
the sbottoms are heavier than $1000 \gev$, and 
in the lower right plot only the points are shown for which additionally
also the sleptons and charginos are heavier than $500 \gev$.
The red line indicates the SM prediction for $M_W$.} 
\label{fig:mstopmw}
\end{figure}
\begin{figure}
\centering
\includegraphics[width=0.48\columnwidth]{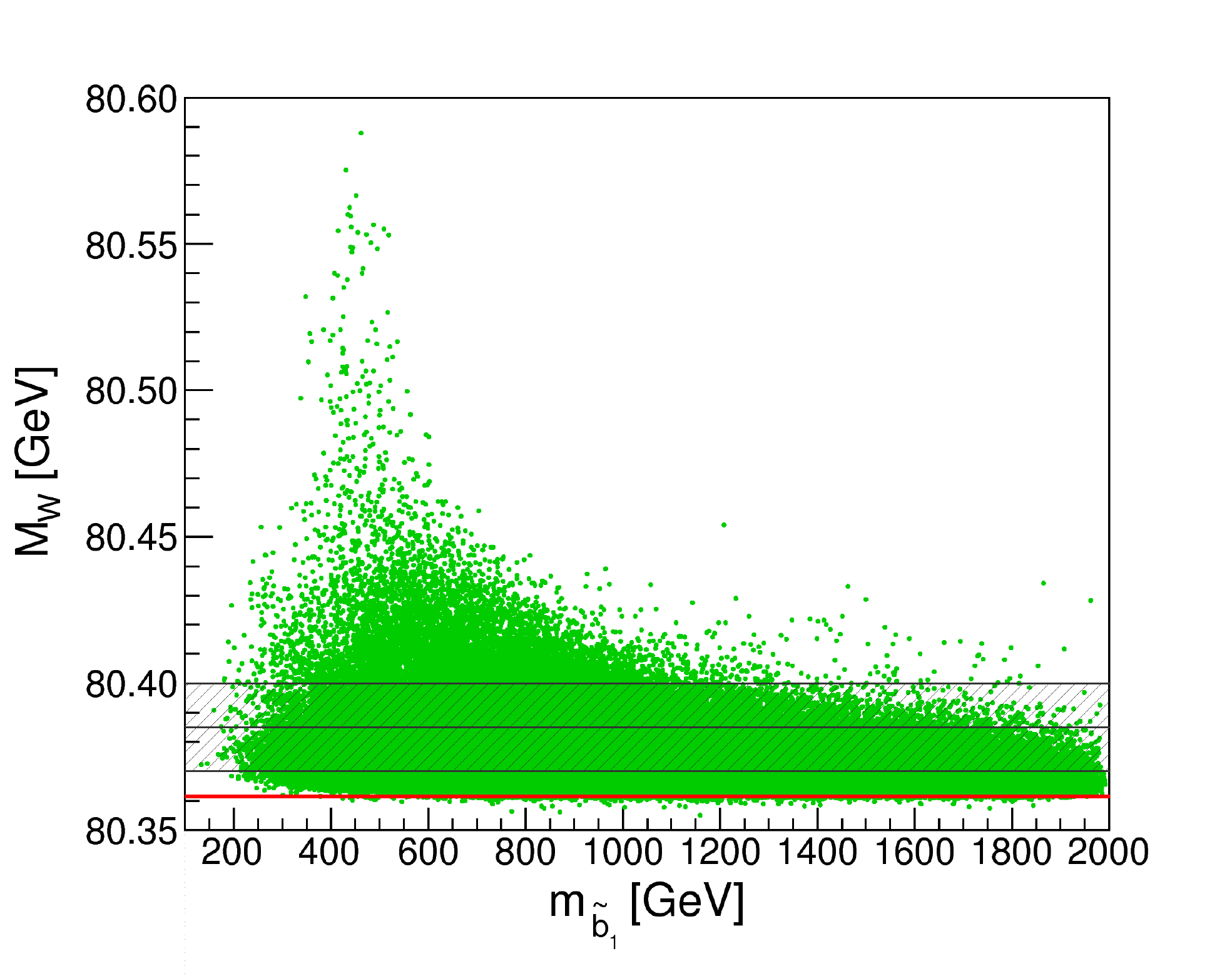}
\includegraphics[width=0.48\columnwidth]{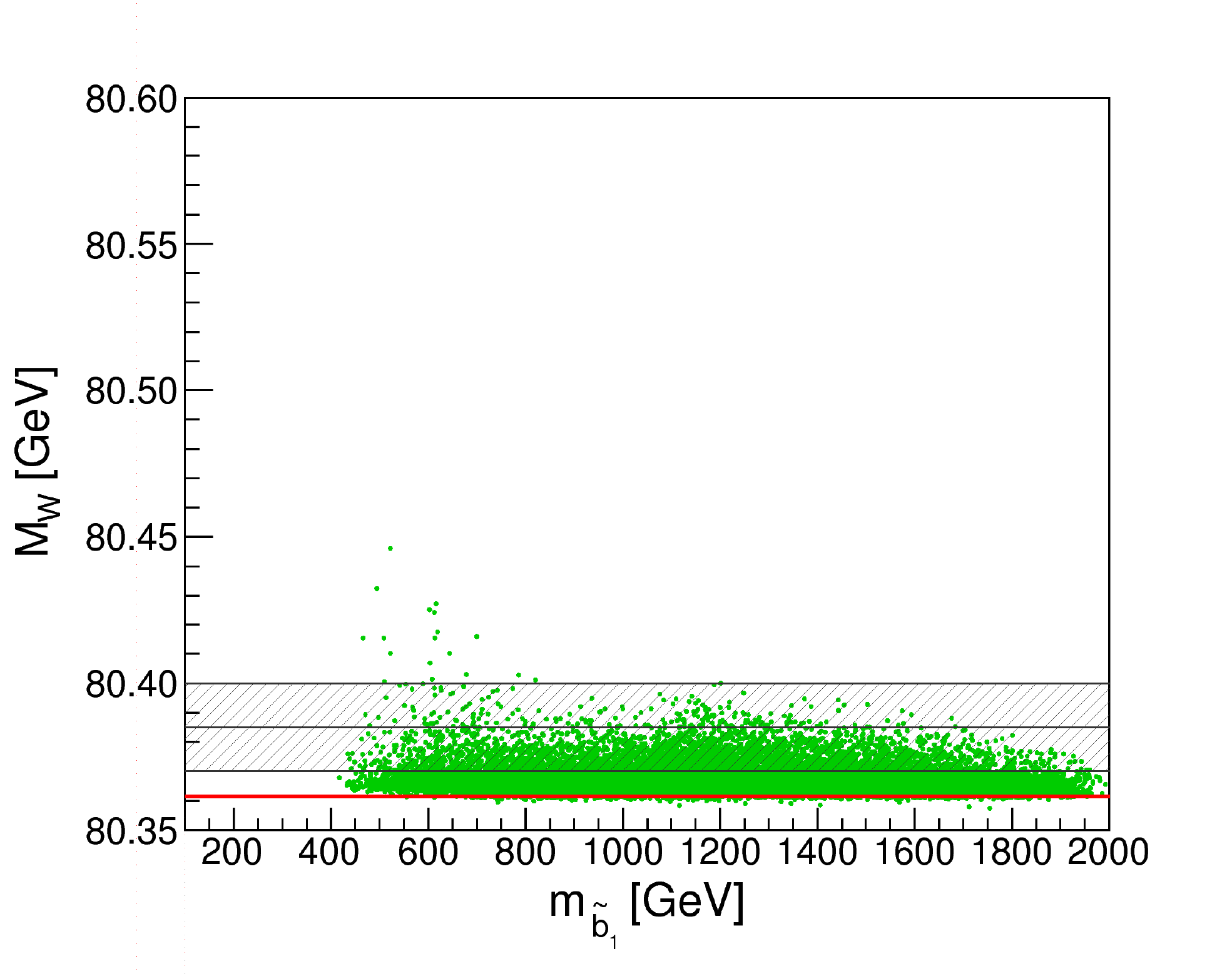}
\caption{Prediction for $\MW$ as a function of the lightest sbottom mass. 
The cuts $m_{\tilde{t}_2}/m_{\tilde{t}_1}<2.5$ and $m_{\tilde{b}_2}/m_{\tilde{b}_1}<2.5$ are applied.
In the left plot all \HiggsBounds\ allowed points are shown, 
in the right plot only the points are shown for which additionally
the squarks of the first two generations and the gluino are heavier
than $1200 \gev$, stops are heavier than $1000 \gev$ and also the sleptons and charginos are heavier than $500 \gev$. As above, the red line indicates the SM prediction for $M_W$.} 
\label{fig:msbottommw}
\end{figure}

\medskip
In \reffi{fig:mstopmw} and \reffi{fig:msbottommw} we analyze in detail the dependence of $\MW$ on
the scalar quark masses, in particular on $\mste$ and $\msbe$, with $\mt$ fixed to $173.2 \gev$. The upper left plot of \reffi{fig:mstopmw} shows the prediction
for $\MW$ (green dots) as a function of $\mste$. All points are allowed
by the constraints discussed in \refse{sec:constraints} and fulfill the
additional constraint
$m_{\tilde{t}_2,\tilde{b}_2}/m_{\tilde{t}_1,\tilde{b}_1}<2.5$. The SM
prediction is shown as a red strip for $\MHSM = 125.6 \pm 0.7 \gev$, 
and the $1\,\si$ experimental result
is indicated as a gray dashed band.
We checked that without the cut
$m_{\tilde{t}_2,\tilde{b}_2}/m_{\tilde{t}_1,\tilde{b}_1}<2.5$ the
largest $\MW$ values are reached for very light stop masses with a very
large ($>2.5$) splitting in the stop sector. Now the maximum of $\sim
80.6 \gev$ is reached for $\mste$ around $800 \gev$. The position where
the maximum is reached depends strongly on the splitting between stops
and sbottoms and will be further explained below (in the discussion of
\reffi{fig:msbottommw}).
In the upper right plot we only show points which
have first and second generation squark masses and the gluino mass 
above $1.2 \tev$, i.e.\ roughly at the limit obtained at the LHC for
simplified spectra~\cite{SUSYsusy1,SUSYsusy2,AtlasSusy,CMSSusy}.
It can be observed that the effects on $\MW$ of the
first and second generation squarks as well as of the gluino are rather
mild. Next, in the lower left plot we only show points which in
addition have $\Sbot$~masses above $1000 \gev$ (this is a
hypothetical cut that is applied for illustration purposes only; it does
not reflect the current experimental situation). 
The fact that all MSSM points in the lower left and lower right plots have stop masses larger than $400 \gev$
results from the restrictions that we have imposed, constraining the sbottom masses ($> 1000 \gev$)
and the maximal splitting in the stop and
sbottom sector ($m_{\tilde{t}_2,\tilde{b}_2}/m_{\tilde{t}_1,\tilde{b}_1}<2.5$) at the same time.
Clearly the sbottoms have a large impact on the $\MW$ prediction. After
applying (for illustration)
the sbottom mass cut the maximal $\MW$ values obtained in the scan are 
$\sim 80.43 \gev$, i.e.\ the SUSY contributions can still be so large in
this case that they can yield not only predicted $\MW$ values that are in good
agreement with the experimental result but also ones that are
significantly higher. The SUSY shift in this case is caused by the
remaining contribution from the
stop--sbottom sector, as well as by the contributions from charginos,
neutralinos and sleptons. In order to disentangle these effects, 
in the lower right plot we also require (again, for illustrative
purposes only) the electroweak SUSY particles to be
heavy and show only points with slepton and chargino masses above $500 \gev$.
A direct mass limit on neutralinos is not applied. Since we fixed 
$M_1 \approx \edz M_2$,
all points have neutralino masses above $\sim 240 \gev$.
In this plot the shift in the $\MW$ prediction as compared to the 
SM case arises solely 
from the stop--sbottom sector with $\msbe > 1000 \gev$ (neglecting the
numerically insignificant contributions from the other sectors for large
SUSY particle masses).
One can observe that $\MW$ values up to the upper edge of the
experimental $1\,\si$ band ($\sim 80.400\gev$) can still be reached for
$\mste$ values as high as $\mste \sim 1100 \gev$ in this case.
For large stop masses, $\mste \gtrsim 1100 \gev$, the contributions 
from the stop--sbottom sector decrease as expected in the decoupling
limit.%
\footnote{In all plots in \reffi{fig:mstopmw} one can see a small gap
between the MSSM points for $\mste > 1900 \gev$ and the SM line. This is
an artefact of the chosen scan ranges: in this region
the mass-splitting between $\Stope$ and $\Stopz$ is small,
and $\mh$ does not reach values up to $\sim 126\gev$. The $\MW$ value
approached in the decoupling limit therefore corresponds to the SM
prediction for a lower Higgs mass value.}

Now we turn to \reffi{fig:msbottommw} showing the $\MW$ prediction
plotted against $\msbe$. 
In the left plot we show all points that are allowed by
\HiggsBounds\ and the other constraints described above (in particular, 
$m_{\tilde{t}_2}/m_{\tilde{t}_1}<2.5$ and $m_{\tilde{b}_2}/m_{\tilde{b}_1}<2.5$
is required). In the right plot only those points are displayed
for which the stops are heavier than $1000 \gev$, the first and
second generation squark masses as well as the gluino mass are above
$1200 \gev$, and the sleptons and charginos are heavier than $500
\gev$. Focusing first on the left plot, one can see that it displays the 
same qualitative features as the upper
left plot of \reffi{fig:mstopmw}. While one would normally expect that
the highest values for $\MW$ are obtained for the smallest values of
$\mste$ and $\msbe$, in the corresponding plots of 
\reffi{fig:mstopmw} and \reffi{fig:msbottommw}
the highest $\MW$ values are found for $\mste \sim 800 \gev$ and 
$\msbe \sim 400 \gev$. This feature is related to the imposed
restriction that the maximal mass splitting for stop and sbottom masses
is limited to be smaller than 2.5. The largest correction to $\MW$
originates from the stop--sbottom contributions to $\Delta \rho$, 
which depend sensitively on the mass splittings between the four squarks
of the third generation. After imposing the limit on the maximal
mass splittings of stops and sbottoms, these contributions become
largest if the relative size of the sbottom mixing,
$|X_b /\mbox{max}(M_{\tilde{Q}_{3}},M_{\tilde{D}_{3}})|$,
reaches its maximum. 
This is realized in this case for $\msbe \sim 400 \gev$ and
$\msbz / \msbe \sim 2.5$, $\mste / \msbe \sim 2$, giving rise to the
maximum around $\mste \sim 800 \gev$ and $\msbe \sim 400 \gev$ in the
upper left plot of \reffi{fig:mstopmw} and the left plot of
\reffi{fig:msbottommw}, respectively. As expected, for higher values of
$\msbe$ the maximum value reached for $\MW$ in \reffi{fig:msbottommw}
decreases, but $\MW$ values as high as the
experimental central value are seen to be possible all the way up to 
$\msbe \sim 2\tev$.
In the right plot the other SUSY particles are required to be rather
heavy (in particular, the stop masses are assumed to be above $1000
\gev$; the other masses are restricted as described above), so that the
impact of the contributions from the sbottom sector becomes apparent.
While rather large contributions are possible for sbottom masses below
about $800\gev$, for the highest values of $\msbe$ shown in the figure
the MSSM prediction for $\MW$ approaches the one in the SM.

\begin{figure}
\centering
\begin{tikzpicture}
    \node[anchor=south west,inner sep=0] at (0,0) {
   \includegraphics[width=0.48\columnwidth]{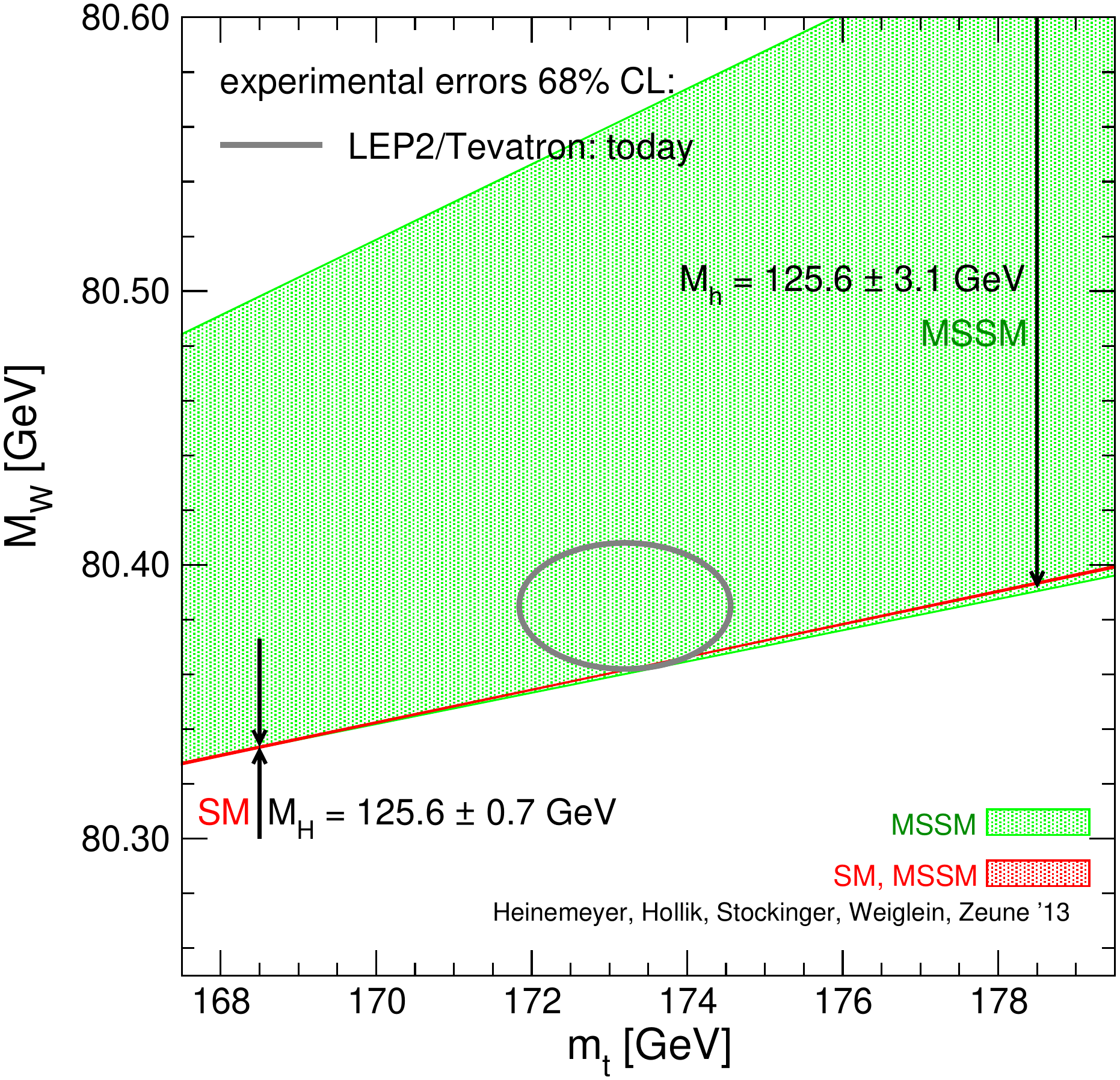}
   \includegraphics[width=0.48\columnwidth]{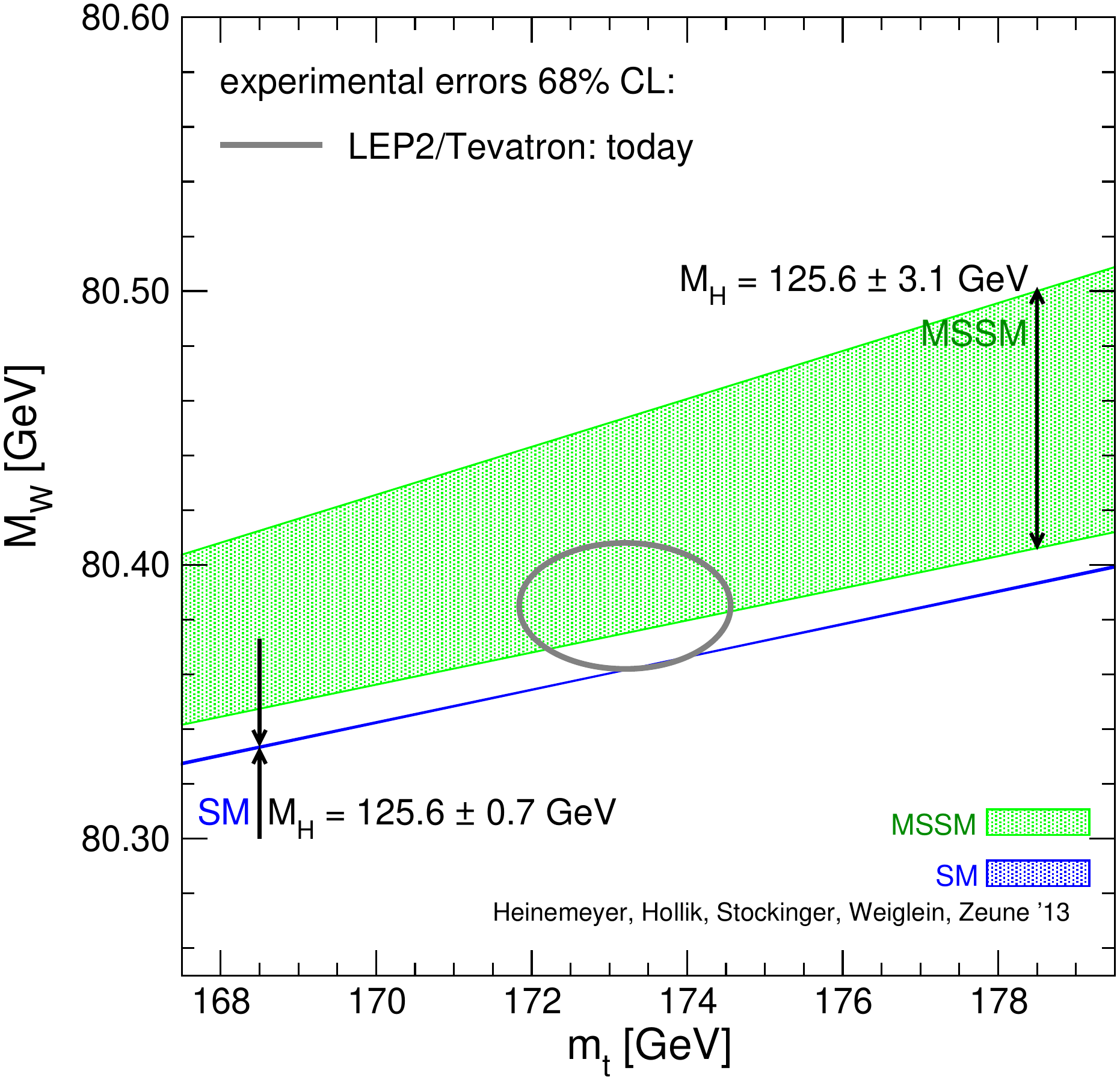}};
    \draw[fill=white,draw=white] (3.4,1) rectangle (7.8,1.3);
    \draw[fill=white,draw=white] (11.3,1.) rectangle (15.8,1.3);
\end{tikzpicture}
\caption{Prediction for $\MW$ as a function of $\mt$.
The left plot shows the $\MW$ prediction assuming the {\em light}
$\cp$-even Higgs boson $h$ in the mass region $125.6 \pm 3.1 \gev$. 
The red band indicates the overlap region of the SM and the MSSM 
with $\MHSM = 125.6 \pm 0.7 \gev$.
The right plot shows the $\MW$ prediction assuming the {\em heavy} $\,\cp$-even
Higgs boson $H$ in the mass region $125.6 \pm 3.1 \gev$.  
The blue band again indicates the SM region with $\MHSM = 125.6 \pm 0.7 \gev$.
All points are allowed by \HiggsBounds.
} 
\label{fig:mtmwmh125}
\end{figure}

\medskip

So far we have only taken into account the existing {\em limits} from the
Higgs searches at the LHC and other colliders (via the program 
\HiggsBounds), but we have not explicitly imposed a constraint in view
of the observed {\em signal} at $\sim 125.6 \gev$. Within the MSSM (referring
to the $\cp$-conserving case for simplicity), the signal can, at least
in principle, be identified either with the light $\cp$-even Higgs boson
$h$ or the heavy $\cp$-even Higgs boson $H$.
In \reffi{fig:mtmwmh125} we show the SM and MSSM prediction of $\MW$ as
a function of $\mt$ as obtained from our scan according to
\refta{tab:scanparam}, where in the left plot the green MSSM area
fulfills $\Mh = 125.6 \pm 3.1 \gev$, while in the right plot the green
MSSM area fulfills $\MH = 125.6 \pm 3.1 \gev$. The substantially larger
uncertainty with respect to the SM experimental uncertainty of $0.7 \gev$ 
(at the $2\,\si$ level) arises as a consequence of the 
theoretical uncertainties from unknown higher-order
corrections in the MSSM prediction for the Higgs boson mass. 
We have added a global uncertainty of
$3 \gev$~\cite{Degrassi:2002fi} in quadrature, yielding a total
uncertainty of $3.1 \gev$. 

Starting with the left plot, where the light $\cp$-even Higgs boson has
a mass that is compatible with the observed signal, we find a similar
result as in \reffi{fig:mtmwFull}. In particular, the comparison with the
experimental results for $\MW$ and $\mt$, indicated by the gray ellipse, 
shows a slight preference for a non-zero SUSY contribution to $\MW$.
While the width of the MSSM area
shown in green is somewhat reduced compared to \reffi{fig:mtmwFull}
because of the additional constraint applied here 
(requiring $\Mh$ to be in the range $\Mh = 125.6 \pm 3.1 \gev$ leads to
a constraint on the stop
sector parameters, see, e.g., \citere{Heinemeyer:2011aa}, which in
turn limits the maximal contribution to $\MW$), the qualitative 
features are the same as in \reffi{fig:mtmwFull}. This is not
surprising, since the limits from the Higgs searches implemented in
\reffi{fig:mtmwFull} have already led to a restriction of the allowed
mass range to the unexcluded region near the observed signal. As in 
\reffi{fig:mtmwFull} the plot shows a small MSSM region (green) below
the overlap region between the MSSM and the SM (red), which is a
consequence of the broadening of the allowed range of $\Mh$ caused by
the theoretical uncertainties from unknown higher-order
corrections, as explained above.

In the right plot of \reffi{fig:mtmwmh125} we show the result for the
case where instead the mass of the heavy $\cp$-even Higgs boson is
assumed to be compatible with the observed signal, i.e.\ 
$\MH = 125.6 \pm 3.1 \gev$. While as mentioned above the interpretation
of the discovered signal in terms of the heavy $\cp$-even Higgs
boson within the MSSM is challenged in particular by the recent ATLAS
bound on light charged Higgs bosons~\cite{AtlaschargedhiggsSUSY13}
(which is not yet included in the version of \HiggsBounds\ used for our
analysis),%
\footnote{If the Higgs sector contains an additional singlet, as in the
NMSSM, it is possible to have a SM-like second-lightest Higgs, while the
charged Higgs boson can be much heavier in this case, see e.g.\
\citere{Stal:2011cz}.}
it is nevertheless interesting to investigate to what extent
the precision observable $\MW$ is sensitive to such a rather exotic scenario
where all five states of the MSSM Higgs sector are light.
The lightest $\cp$-even Higgs in this scenario has a heavily suppressed
coupling to gauge bosons and a mass that can be significantly below the
LEP limit for a SM-like Higgs, see e.g.\ \citere{Carena:2013qia}. As
shown in the right plot of \reffi{fig:mtmwmh125}, the constraint 
$\MH = 125.6 \pm 3.1 \gev$ gives rise to a situation where the MSSM
region (green) does not overlap with the SM prediction (blue). This gap
between the predictions of the two models is caused by the fact that 
$\MH = 125.6 \pm 3.1 \gev$ implies
light states in the Higgs sector (in particular a light charged Higgs), which lead to a non-zero SUSY 
contribution to $\MW$ in this case, 
whereas for the light $\cp$-even
Higgs boson the constraint $\Mh = 125.6 \pm 3.1 \gev$ can be fulfilled
in the decoupling region of the MSSM. The plot furthermore shows that 
the constraint $\MH = 125.6 \pm 3.1 \gev$ implies not only a lower bound
on the SUSY contribution to $\MW$ but also a more restrictive upper
bound, as can be seen from comparing the two plots in
\reffi{fig:mtmwmh125}.
It is
interesting to note that also in the case where the heavy $\cp$-even
Higgs is in the mass range compatible with the observed signal, 
the MSSM turns out to be
better compatible with the experimental results for $\MW$ and $\mt$
(indicated by the gray ellipse) than the SM.

\begin{figure}
\centering
\includegraphics[width=0.68\columnwidth]{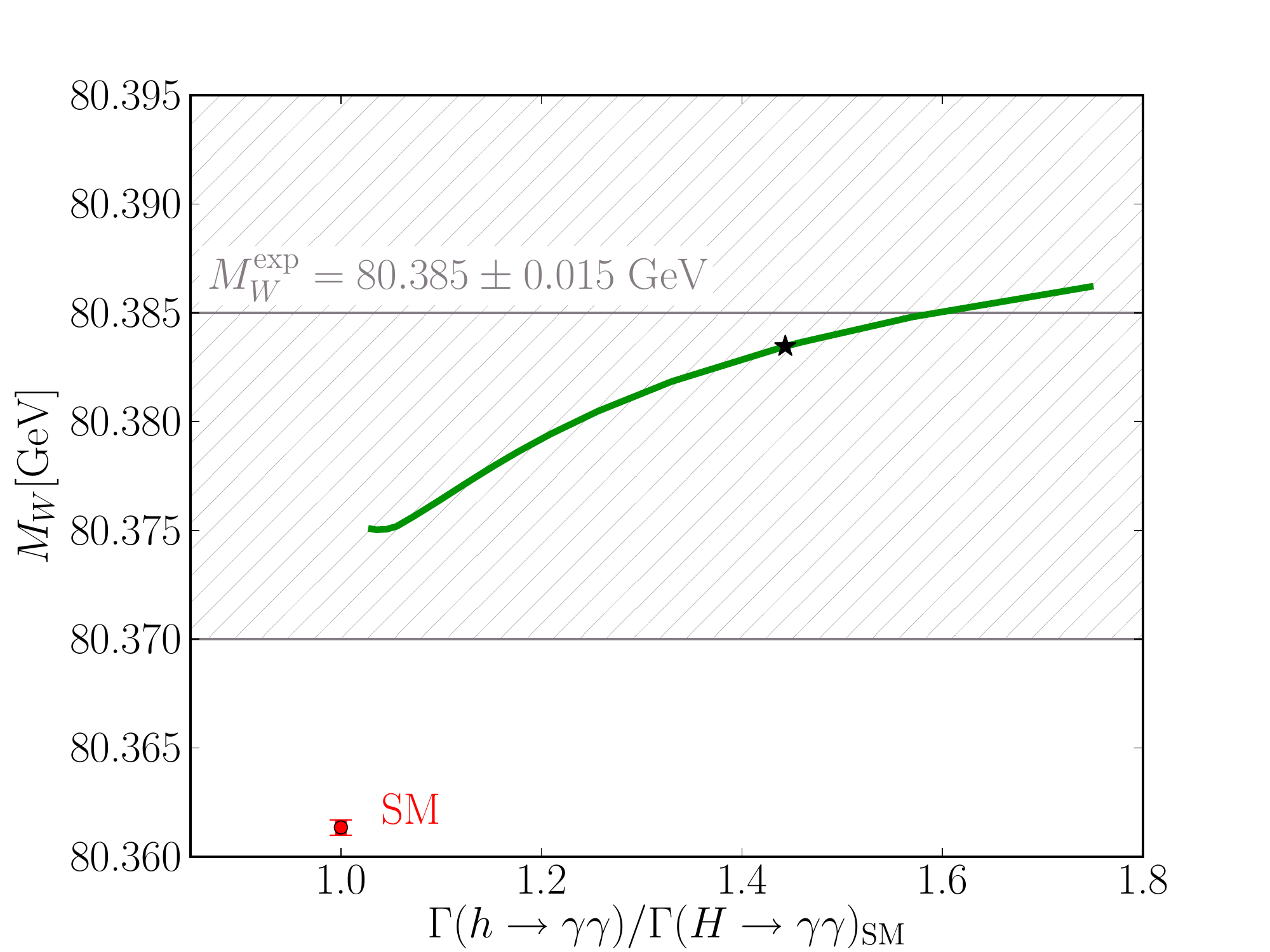}
\caption{$\MW$ prediction in the MSSM as a function of 
  $\Gamma (h \to \gamma \gamma)$, normalized to the SM value. 
  The black star indicates
  the best fit point from a pMSSM-7 fit to
  all Higgs data (available at that time)~\cite{Bechtle:2012jw}.
  The green line is obtained by varying
  $M_{\tilde{E}_{3}}=M_{\tilde{L}_{3}}$ from $280 \gev$ to $500 \gev$.
} 
\label{fig:mstaumw}
\end{figure}

\medskip
In \reffi{fig:mstaumw} we analyze the dependence of the $\MW$ prediction
on light scalar taus. 
In \citeres{Carena:2012gp,Carena:2011aa} it was shown
that light scalar taus can 
enhance the decay rate of the light $\cp$-even Higgs boson into photons.
This is of interest in view of the current experimental situation,
where the signal strength in the $\ga\ga$ channel observed by 
ATLAS~\cite{ATLASgagaSummer13} lies
significantly above the value expected in the SM (but is 
still compatible at the $2\,\sigma$ level), while the signal strength
observed in CMS~\cite{CMSHiggsgagaSummer13} 
is currently slightly below the SM level.
Since loop contributions of BSM particles to the decay width 
$\Ga(h \to \ga\ga)$ do not have to compete with a SM-type tree-level
contribution, this loop-induced quantity is of particular relevance for
investigating possible deviations from the SM prediction.
\reffi{fig:mstaumw} shows
the prediction for
$\MW$ as a function of $\Ga(h \to \ga\ga)/\Ga(H \to \ga\ga)_{\SM}$,
where the latter has been evaluated with \fh. As a starting point we use
the best-fit point obtained in \citere{Bechtle:2012jw} from a pMSSM-7 fit to
all Higgs data (available at that time), which indeed exhibited an
enhancement of $\Ga(h \to \ga\ga)$ due to scalar taus with a mass close to
$100 \gev$. The parameters of the best fit point are
$\MA =669 \gev$,  $\tb=16.5$, $\mu=2640 \gev$,
$M_{\tilde{Q}_3}=M_{\tilde{U}_3}=M_{\tilde{D}_3}=1100 \gev$, 
$M_{\tilde{Q}_{1,2}}=M_{\tilde{U}_{1,2}}=M_{\tilde{D}_{1,2}}=1000 \gev$, 
$M_{\tilde{L}_3}=M_{\tilde{E}_3}=285 \gev$, 
$M_{\tilde{L}_{1,2}}=M_{\tilde{E}_{1,2}}=300 \gev$, $A_f = 2569 \gev$, 
$M_2 = 201 \gev$ and $M_3=1000 \gev$.
In \reffi{fig:mstaumw} the best-fit point
is indicated as a black star.
We vary the stau mass scale $M_{\tilde E_3} = M_{\tilde L_3}$ 
in the range of $280 \gev$ to $500 \gev$, giving rise to
a corresponding variation of the lighter stau mass.
The results are shown as the green line in
\reffi{fig:mstaumw}, where the current experimental 
$1\,\si$ region for $\MW$
is indicated as a gray band. One can observe that for light scalar taus, 
corresponding to larger $\Ga(h \to \ga\ga)$, the
agreement of the prediction for $\MW$ with the experimental value is 
improved. 
A certain level of enhancement of $\Ga(h \to \ga\ga)$ is also compatible
with the current experimental results on the signal strength in the
$\ga\ga$ channel.
For heavy scalar taus, as obtained for 
$M_{\tilde E_3} = M_{\tilde L_3} = 500 \gev$ (and keeping the other
parameters as defined above),
the $\MW$ prediction still
remains within the experimental $1\,\si$ band, while nearly SM values for
$\Ga(h \to \ga\ga)$ are reached. 

\subsection{Discussion of possible future scenarios}
\label{sec:future}

In the final step of our investigation we discuss the precision
observable $\MW$ in the context of possible {\em future} scenarios.
We first investigate the impact of an assumed limit of $500 \gev$ on stops
and sbottoms (and assume that no other colored particles are observed
below $1200 \gev$).
\begin{figure}
\centering
\begin{tikzpicture}
    \node[anchor=south west,inner sep=0] at (0,0) {
\includegraphics[width=0.33\columnwidth]{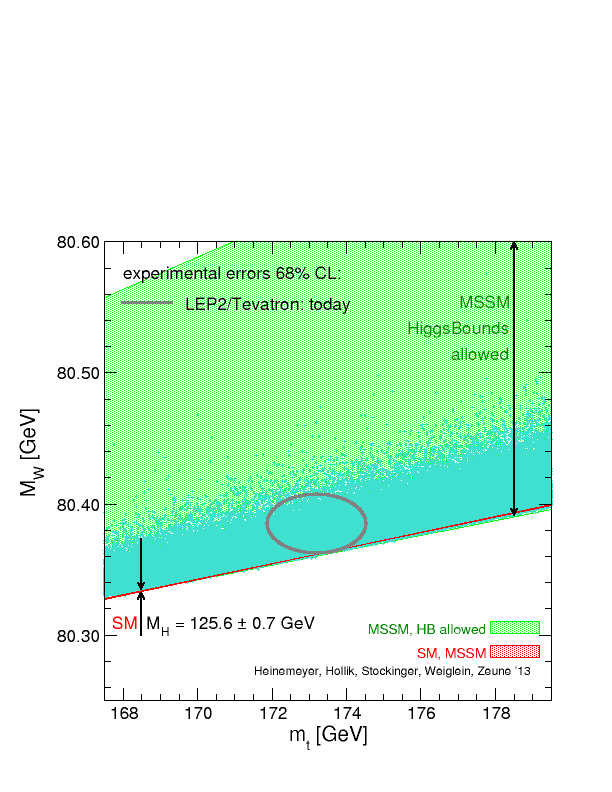}
\includegraphics[width=0.33\columnwidth]{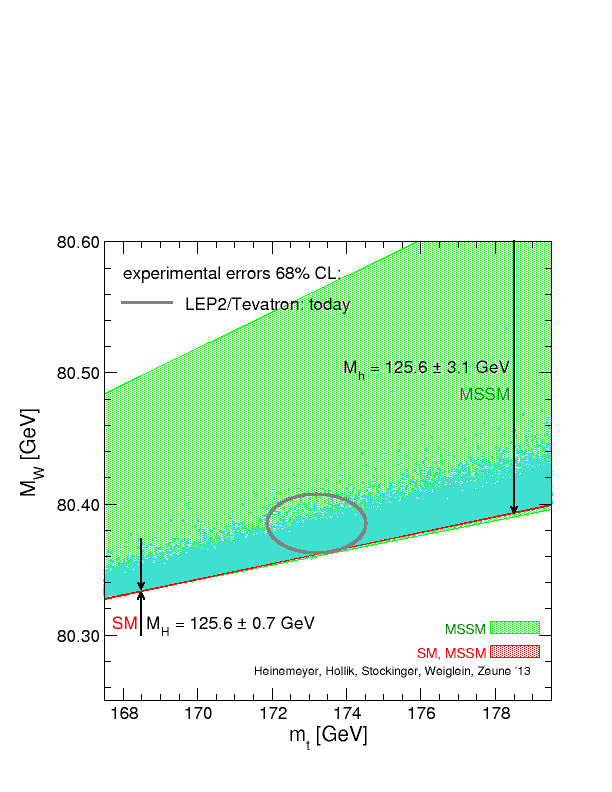}
\includegraphics[width=0.33\columnwidth]{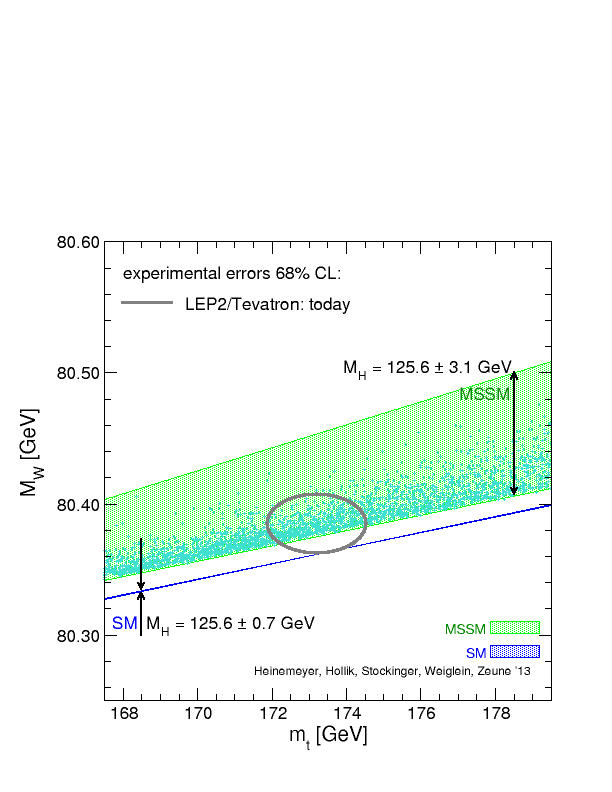}};
   \draw[fill=white,draw=white] (2.2,1.00) rectangle (4.8,1.15);
   \draw[fill=white,draw=white] (7.7,1.00) rectangle (10.35,1.15);
   \draw[fill=white,draw=white] (13.1,1.00) rectangle (16,1.15);
\end{tikzpicture}
\caption{Prediction for $\MW$ as a function of $\mt$.
  The left
  plot shows all points allowed by \HiggsBounds, the middle one
  requires $\Mh$ to be in the mass region $125.6 \pm 3.1 \gev$, while
  in the right plot $\MH$ is required to be in
  the mass region $125.6 \pm 3.1 \gev$. 
  The color coding is as in \reffis{fig:mtmwFull} and 
  \ref{fig:mtmwmh125}.
  In addition, the blue points are the parameter
  points for which the stops and sbottoms are heavier than $500 \gev$ and
  squarks of the first two generations and the gluino are heavier than
  $1200 \gev$.} 
\label{fig:mtmwmh125mstop}
\end{figure}
In \reffi{fig:mtmwmh125mstop} we show again the $\MW$--$\mt$ planes as
presented in \reffi{fig:mtmwFull} (where the parameter region
allowed by \HiggsBounds\ is displayed) and in
\reffi{fig:mtmwmh125} ($\Mh$ or $\MH$ in the range of $125.6 \pm 3.1 \gev$),
but now in addition the light blue points obey the (hypothetical) mass
limits for stops and sbottoms ($500 \gev$) and for other colored particles
($1200 \gev$).
The left plot shows the \HiggsBounds\ allowed points,
whereas in the middle (right) plot $\Mh (\MH) = 125.6 \pm 3.1 \gev$ 
is required.
It can be observed that the light blue points corresponding to a 
relatively heavy
colored spectrum are found at the lower end of the predicted $\MW$
range, i.e.\ in the decoupling region of the MSSM.
As discussed above the largest SUSY contributions arise from the stop--sbottom
sector. If lower lower mass limits on stops and sbottoms of
$500 \gev$ are assumed, it can be seen that the band corresponding to
the possible range of predictions for $\MW$ in the MSSM 
would shrink significantly, to the
region populated by the blue points.
It should be noted that the prediction for $\MW$ in this region is in
perfect agreement with the experimental measurements of
$\MW$ and $\mt$. 
Besides the contributions of stops and sbottoms, which can still be
significant even if the stops and sbottoms are heavier than $500 \gev$, 
the main SUSY corrections arise
from relatively light sleptons, charginos and neutralinos, as analyzed
above.

\begin{figure}
\centering
\begin{tikzpicture}
    \node[anchor=south west,inner sep=0] at (0,0) {
   \includegraphics[width=0.48\columnwidth]{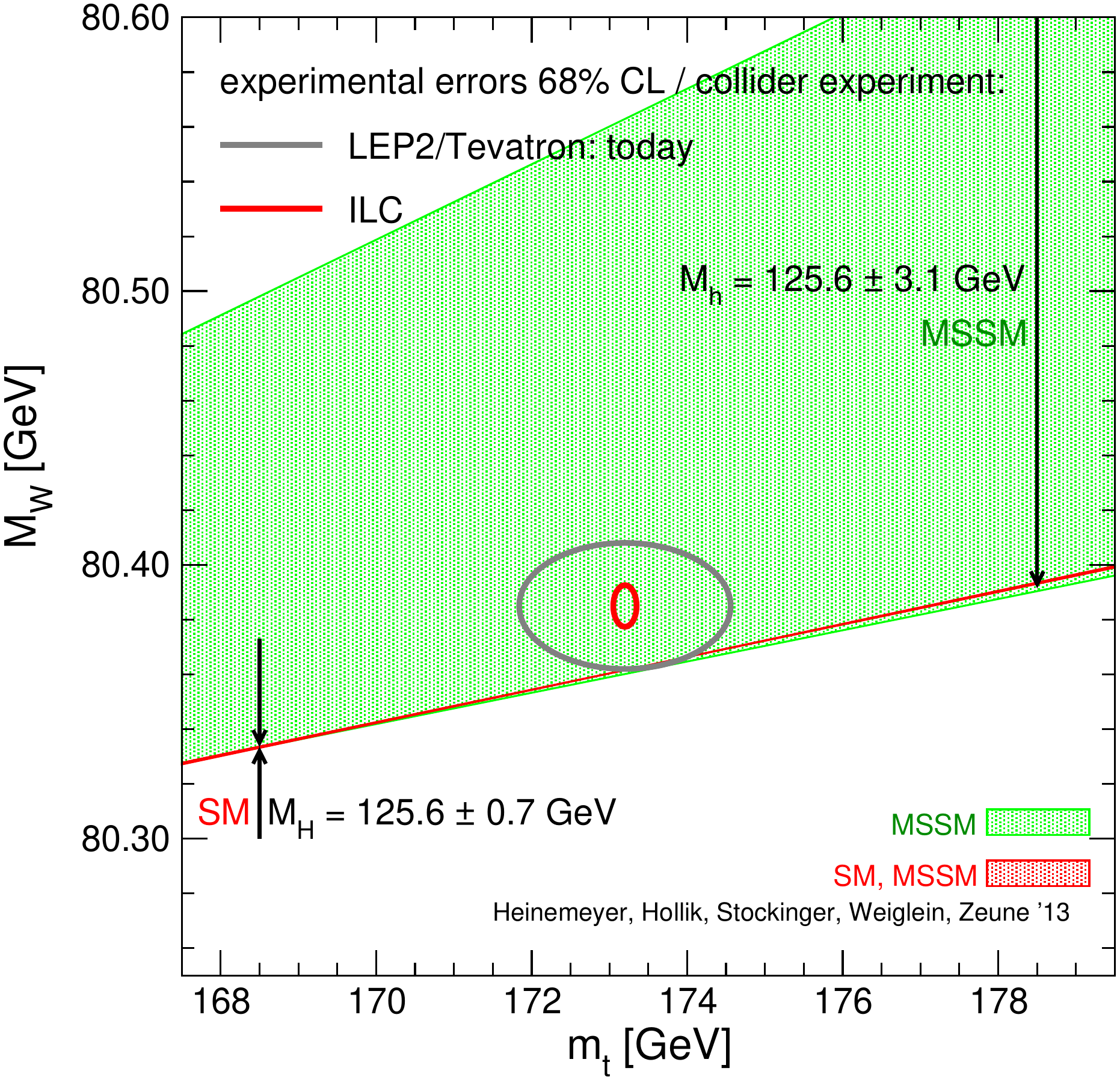}};
  \draw[fill=white,draw=white] (3.4,1) rectangle (7.8,1.3);
\end{tikzpicture}
\caption{Prediction for $\MW$ as a function of $\mt$, 
as given in the left plot of~\reffi{fig:mtmwmh125} (the mass $\Mh$ of the
light $\cp$-even Higgs boson is assumed to be in the region 
$125.6 \pm 3.1 \gev$). 
In addition to the current experimental results for $\MW$ and $\mt$ that
are displayed by the gray 68\%~C.L.\ ellipse the anticipated future
precision at the ILC is indicated by the red ellipse (assuming the same
experimental central values).
} 
\label{fig:mtmwmh125lcprecision}
\end{figure}

\medskip
While so far we have compared the various predictions with the
current experimental results for $\MW$ and $\mt$, we now discuss the
impact of future improvements of these measurements. For the $W$ boson
mass we assume an improvement of a factor three compared to the present
case down to $\Delta \MW= 5 \mev$ from future measurements at the LHC
and a prospective Linear Collider (ILC)~\cite{Baak:2013fwa}, while for $\mt$ we adopt the
anticipated ILC accuracy of $\Delta \mt = 100\mev$~\cite{Baer:2013cma}.
For illustration we show in \reffi{fig:mtmwmh125lcprecision}
again the left plot of
\reffi{fig:mtmwmh125}, assuming the mass of the light
$\cp$-even Higgs boson $h$ in the region $125.6 \pm 3.1 \gev$, but
supplement the gray ellipse indicating the present experimental results
for $\MW$ and $\mt$ with the future projection indicated by the red
ellipse (assuming the same experimental central values).
While currently the experimental results for $\MW$ and $\mt$ are 
compatible with the predictions of both models (with a slight
preference for a non-zero SUSY contribution), the anticipated future
accuracies indicated by the red ellipse would clearly provide a high
sensitivity for discriminating between the models and for constraining
the parameter space of BSM scenarios.

\begin{figure}
\begin{center}
\includegraphics[width=0.48\hsize]{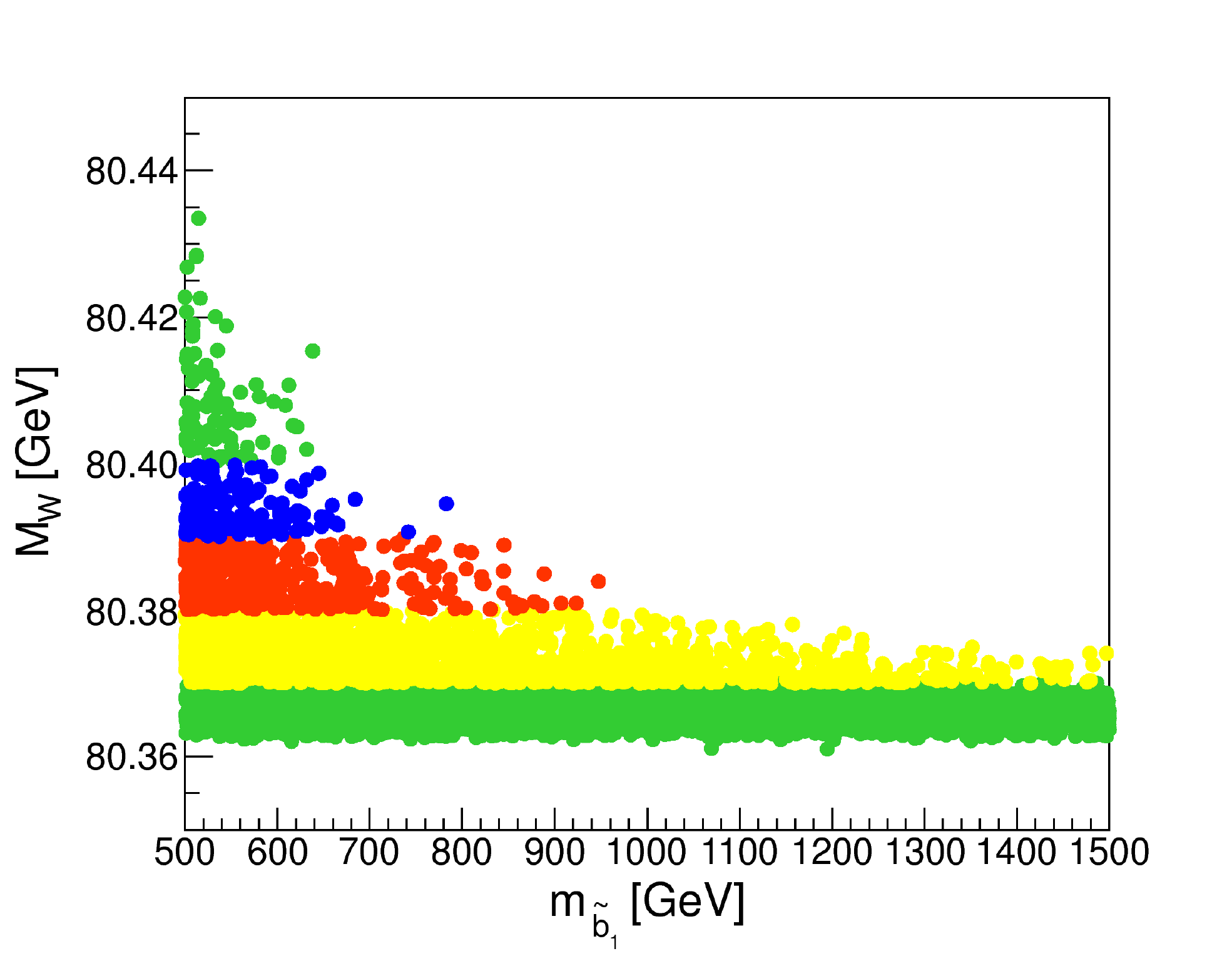}
\includegraphics[width=0.48\hsize]{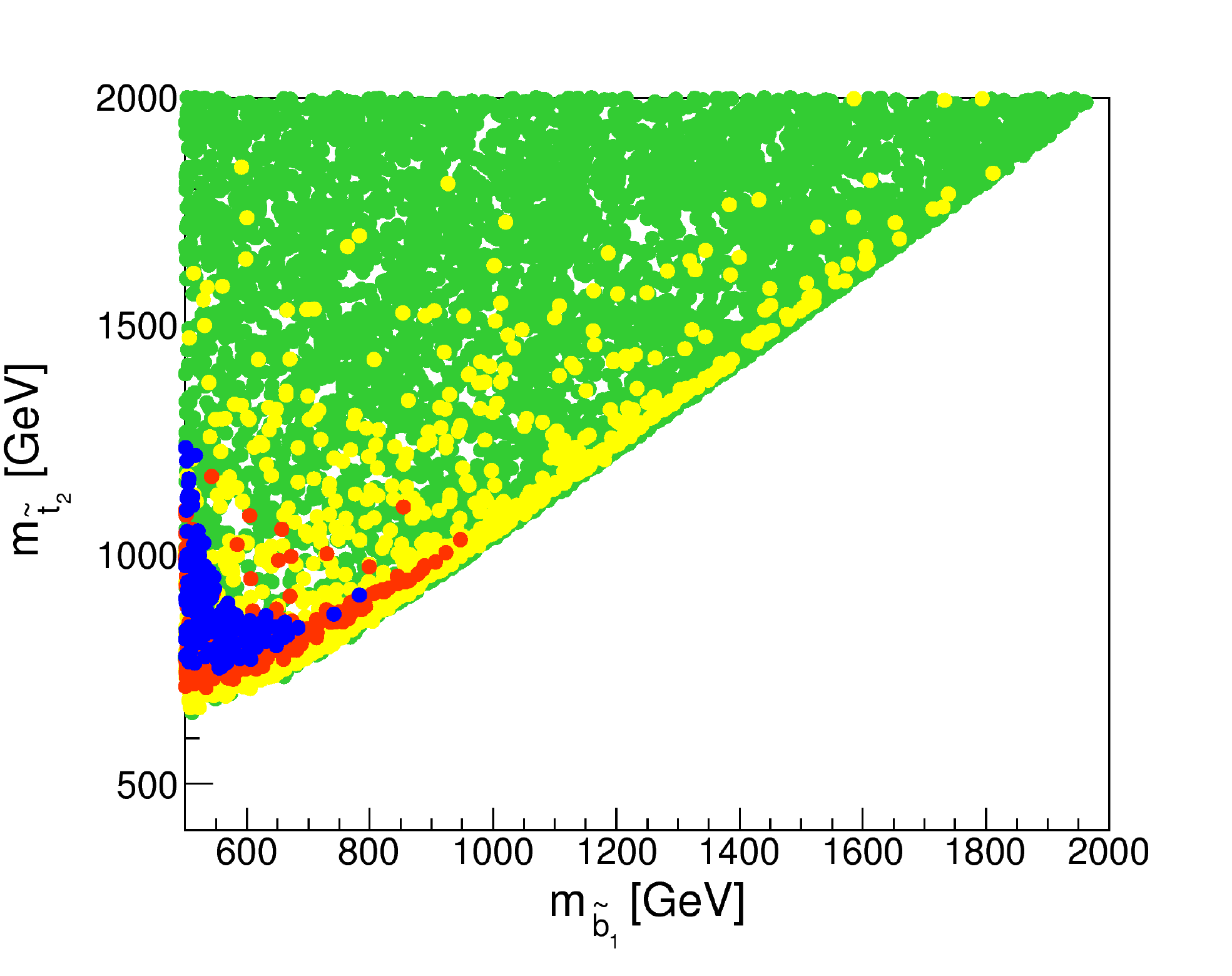}
\caption{Results of an MSSM parameter scan illustrating the 
prediction for $\MW$ in a hypothetical future scenario
assuming a measurement of $\mste = 400 \pm 40 \gev$ at the LHC as well
as lower limits on all other SUSY particles: the assumed lower limits
are $500 \gev$ for the other third generation squarks, $1200 \gev$ for
all other colored particles, and $300 \gev$ for sleptons and charginos.
All displayed points fulfill $\Mh = 125.6 \pm 3.1 \gev$. The yellow,
  red and blue points correspond to $\MW=80.375 \pm 0.005 \gev$ (yellow), 
  $\MW=80.385 \pm 0.005 \gev$ (red), and $\MW=80.395 \pm 0.005 \gev$ (blue).
The left plot shows the prediction for $\MW$ as a function of the
lighter sbottom mass, $\msbe$, while the right plot shows the $\MW$
prediction in the $\msbe$--$\mstz$ plane.}
\label{fig:electroweak-fig1lisa}
\end{center}
\end{figure}

\medskip
As a further hypothetical future scenario we assume that a
light scalar top quark has been discovered at the LHC with a mass of 
$\mste = 400 \pm 40 \gev$, while no other new particle has been
observed. 
As before, for this analysis we use
an anticipated experimental precision of 
$\De\MW = 5\mev$ (other uncertainties have been neglected in this analysis).
Concerning the masses of the other SUSY particles, we assume
lower limits of 
$300 \gev$
on both sleptons and 
charginos, 
$500 \gev$ on other 
scalar quarks of the third generation and of $1200 \gev$ on the
remaining colored particles. We have selected the points from our scan
accordingly. Any additional particle observation would impose a
further constraint and would thus enhance the sensitivity of the
parameter determination.
In \reffi{fig:electroweak-fig1lisa}
we show the 
parameter points from our scan that are compatible with the above
constraints.
All points fulfill $\Mh = 125.6 \pm 3.1 \gev$ and 
$\mste = 400 \pm 40 \gev$. Yellow, red and blue 
points have furthermore a $W$~boson mass of 
$\MW = 80.375, 80.385, 80.395 \pm 0.005 \gev$, respectively, 
corresponding to three hypothetical future central experimental values
for $\MW$.
The left plot in \reffi{fig:electroweak-fig1lisa} shows the $\MW$
prediction as a function of the lighter sbottom mass.  
Assuming that the experimental central value for 
$\MW$ stays at its current value of $80.385 \gev$
(red points) or goes up by $10 \mev$ (blue points),  
the precise measurement of $\MW$ would set stringent upper
limits of $\sim 800 \gev$ (blue) or $\sim 1000 \gev$ (red) on the 
possible mass range of the lighter sbottom.
As expected, this sensitivity degrades if the experimental central value
for $\MW$ goes down by $10 \mev$ (yellow points), which would bring it
closer to the SM value given in \refeq{eq:mwsmresult}.
The right plot shows the results in the $\msbe$--$\mstz$ plane.
It can be observed that sensitive upper bounds on those
unknown particle masses could be set%
\footnote{See also \citere{Barger:2012hr} for a recent analysis
investigating constraints on the scalar top sector.}
based on an experimental value of
$\MW$ of $80.385 \pm 0.005 \gev$ or $80.395 \pm 0.005 \gev$
(i.e.\ for central values sufficiently different from the SM
prediction). In this situation 
the precise $\MW$ measurement could give interesting indications 
regarding the
search for the heavy stop and the light sbottom
(or put the interpretation within the MSSM under tension). 

\section{Conclusions}
\label{sect:conclusions}

We have presented the currently most precise prediction for the
$W$~boson mass in the MSSM and compared it with the state--of--the--art
prediction in the SM. The evaluation in the MSSM includes the
full one-loop result (for the general case of complex parameters) and
all known higher-order corrections of SM and SUSY type. Within the SM,
interpreting the signal discovered at the LHC as the SM Higgs boson with
$\MHSM = 125.6 \gev$,
there is no unknown parameter in the $\MW$ prediction anymore. This
yields $\MW^{\text{SM}}=80.361 \gev$, which is somewhat below (but
compatible at the level of about $1.5\,\si$) with the current experimental
value of $\MW^{\rm exp} = 80.385 \pm 0.015 \gev$. The loop contributions
from supersymmetric particles in general give rise to an upward shift in
the prediction for $\MW$ as compared to the SM case, which tend to bring
the prediction into better agreement with the experimental result. For
very light superpartners of the top and bottom quarks and large mass
splittings in this sector even much larger (and thus experimentally
disfavored) values of $\MW$ are possible. 

We have investigated the MSSM and SM predictions in the $\MW$--$\mt$
plane, updating earlier results in~\citere{Heinemeyer:2006px} while
taking into account the existing constraints from Higgs and SUSY
searches. We have analyzed in this context the implications of the
results of present and possible future searches for supersymmetric
particles at the LHC. While the existing 
bounds on the gluino and the squarks of 
the first two generations have only a minor effect, more stringent
bounds on the third generation squarks would have a drastic effect on
the possible range of $\MW$ values in the MSSM. In particular, assuming
a lower bound of $500\gev$ on the masses of the stops and sbottoms, the 
resulting range of predicted $\MW$ values in the MSSM essentially
reduces to the region that is best compatible with the experimental
result (corresponding to the 68\% C.L.\ region). 
We have shown that MSSM predictions in exact agreement
with the current experimental central value of $\MW$ can be reached for
stop mass values as large as $\mste \sim 1.5 \tev$,
even if all other SUSY particles are heavy. We have furthermore pointed
out that even if the squarks are so heavy that their contribution to
$\MW$ becomes negligible, sizable SUSY
contributions to $\MW$ are nevertheless 
possible if either charginos, neutralinos or sleptons are
light. Analyzing the impact of light SUSY particles that are still
allowed by LHC searches we have found that scalar leptons can give a
contribution larger than
$60 \mev$, while light charginos can give corrections of up to $\sim 20
\mev$.

Besides the impact of limits from searches for supersymmetric particles,
we have analyzed the constraints arising from the Higgs signal
at about $\MHexp \gev$. Within the MSSM this signal can be interpreted,
at least in principle, either as the light or the heavy $\cp$-even Higgs 
boson (we have not addressed here the possibility of a state consisting
of an admixture of $\cp$-even and $\cp$-odd components).
Concerning the interpretation in terms of the light $\cp$-even Higgs
boson, the result for $\MW$ turns out to be well compatible with the 
additional constraint that $\Mh$ should be in the 
mass range compatible with the signal. The main effect of this
constraint is that it somewhat reduces the allowed range of predicted
$\MW$ values in the MSSM, improving in this way the overall
compatibility with the experimental result for $\MW$. It is remarkable
that also the rather exotic scenario where the mass of the heavy
$\cp$-even Higgs boson is required to be in the range compatible
with the observed signal (which is under pressure in particular from the
recent ATLAS bound on light charged Higgs bosons) leads to predicted
values for $\MW$ that tend to be in better agreement with the
experimental result than for the SM case. It is interesting to note that
in this case, which corresponds to an MSSM scenario outside of the
decoupling region, there is no overlap between the SM prediction and the
range of MSSM predictions for $\MW$. A high-precision measurement of
$\MW$ could thus yield a clear distinction between the two models in
such a scenario.

As another interesting feature in the context of Higgs phenomenology, we 
have studied the correlation between $\MW$ and $\Ga(h \to \ga\ga)$ via light
scalar taus. Light staus contribute to the loop-induced process
$h \to \ga\ga$, leading to an enhancement of the $\ga \ga$ width
over the
SM prediction. At the same time staus appear in the MSSM loop
corrections to the muon decay, and thus light staus can also yield a
sizable contribution to the prediction for $\MW$.
We have demonstrated that light staus can have the simultaneous effect
of enhancing $\Ga(h \to \ga\ga)$ while bringing the $\MW$ prediction in
perfect agreement with the current experimental central value of
$\MW$.

As a final step we have discussed the impact of the precision observable 
$\MW$ in the context of
possible future scenarios. The improved precision on $\MW$ and $\mt$
from future measurements at the LHC and in particular at a prospective
Linear Collider (ILC) would significantly enhance the sensitivity to
discriminate between the SM and the MSSM (as well as other BSM
scenarios). Analyzing in this context the impact of possible future LHC 
results in the stop sector on the $\MW$ prediction, we have discussed a
hypothetical scenario where a light stop has been detected at the LHC,
while lower limits have been imposed on all other SUSY particles. We
have demonstrated that, depending on the future central experimental
value, a high-precision measurement of $\MW$ could yield
quite stringent {\em upper\/} bounds on the mass of the heavier stop and
the lighter sbottom, which could be of great interest regarding the
direct searches for those particles. 
In case other SUSY particles were detected, this would further sharpen the
sensitivity for determining unknown mass scales of the model.

\subsection*{Acknowledgements}
We are grateful to 
A.~Freitas, 
T.~Hahn,
A.~Kotwal,
O.~St{\aa}l,
T.~Stefaniak
and
D.~Wackeroth for helpful discussions.
This work has been supported by 
the Collaborative Research Center SFB676 of the DFG,
``Particles, Strings and the early Universe".
The work of S.H.\ was supported in part by CICYT 
(grant FPA 2010--22163-C02-01) and by the Spanish MICINN's 
Consolider-Ingenio 2010 Program under grant MultiDark CSD2009-00064.


\bibliography{ref}
\bibliographystyle{JHEP}


\end{document}